\documentclass{aa}

\usepackage{natbib}
\bibpunct{(}{)}{;}{a}{}{,} 

\usepackage{lscape}

\usepackage{graphicx}
\usepackage{amsmath}
\usepackage{txfonts}
\graphicspath{{Figures/AllFigures/}}

\newcommand{\newtext}[1]{#1}
\usepackage{color}
\usepackage{subcaption}

\begin{document} 

   \title{HD far infrared emission as a measure of protoplanetary disk mass}

   \author{L. Trapman \inst{1}
          \and
          A. Miotello \inst{1}
          \and 
          M. Kama \inst{1}
          \and 
          E.F. van Dishoeck \inst{1,2}
          \and 
          S. Bruderer \inst{2}
          }

   \institute{
            Leiden Observatory, Leiden University, Niels Bohrweg 2, NL-2333 CA Leiden, The Netherlands \\
            \email{trapman@strw.leidenuniv.nl}
        \and
            Max-Planck-institute f\"{u}r extraterrestrische Physic, Giessenbachstra{\ss}e, D-85748 Garching, Germany
             }
   \date{Received 21 December 2016; accepted 17 May 2017}

 
  \abstract
   {Protoplanetary disks around young stars are the sites of planet formation. While the dust mass can be estimated using standard methods, determining the gas mass - and thus the amount of material
available to form giant planets - has proven to be very difficult. Hydrogen deuteride (HD) is a promising alternative to the commonly-used gas mass tracer, carbon monoxide. However, its potential has not yet been investigated with models incorporating both HD and CO isotopologue-specific chemistry, and its sensitivity to uncertainties in disk parameters has not yet been quantified.}
   {To examine the robustness of HD as tracer of the disk gas mass, specifically the effect of gas mass on the HD far infrared emission and its sensitivity to the vertical structure. Also, to provide requirements for future far-infrared missions such as SPICA.}
   {Deuterium chemistry reactions relevant for HD were implemented in the thermochemical code \texttt{DALI} and more than 160 disk models were run for a range of disk masses and vertical structures.}
   {The HD J=1-0 line intensity depends directly on the gas mass through a sublinear power law relation with a slope of $\sim$0.8. Assuming no prior knowledge about the vertical structure of a disk and using only the HD 1-0 flux, gas masses can be estimated to within a factor of two for low mass disks (M$_{\rm disk} \leq 10^{-3}$ M$_{\odot}$). For more massive disks, this uncertainty increases to more than an order of magnitude. Adding the HD 2-1 line or independent information about the vertical structure can reduce this uncertainty to a factor of $\sim3$ for all disk masses. For TW Hya, using the radial and vertical structure from \cite{Kama2016} the observations constrain the gas mass to $6\cdot10^{-3}\ \mathrm{M}_{\odot}\leq\mathrm{M}_{\rm disk}\leq9\cdot10^{-3}$ M$_{\odot}$. Future observations require a $5\sigma$ sensitivity of $1.8\cdot10^{-20}$ W m$^{-2}$ ($2.5\cdot10^{-20}$ W m$^{-2}$) and a spectral resolving power $R\geq300\ (1000)$ to detect HD 1-0 (HD 2-1)  for all disk masses above $10^{-5}\ \mathrm{M}_{\odot}$ with a line-to-continuum ratio $\geq0.01$.} 
   { These results show that HD can be used as an independent gas mass tracer with a relatively low uncertainty and should be considered as an important science goal for future far-infrared missions.}


   \maketitle
%

\section{Introduction}
\label{sec: introduction}

One of the key properties for understanding the evolutionary path of disks and the formation of planets is the disk gas mass (e.g., \citealt{Armitage2015}). Determining gas masses is no easy task. The main component of the gas, molecular hydrogen (H$_2$), is difficult to observe. As a symmetric rotor, H$_2$ has no electric dipole moment, limiting its rotational lines to quadrupole transitions ($\Delta J = 2$). The $J = 2$ level lies at $549.2$ K, so in order to produce any appreciable H$_2$ emission gas temperatures have to be at least 100 K \citep{Thi2001,Carmona2011} which is much higher than the temperature of the bulk of the gas in the disk. In addition, these lines lie in the mid- and near-infrared (IR) where the continuum is bright, which further hampers the observations due to the high continuum optical depth and low line-to-continuum ratios. Even if detected, the H$_2$ emission does not trace the bulk of the mass (e.g., \citealt{Pascucci2013}). This means that indirect tracers of the gas mass have to be used. 

Historically, most of the disk mass estimates have been based on observations of the (sub-)millimeter (mm) emission of the dust grains (e.g., \citealt{Beckwith1990, Dutrey1996,ManningsSargent1997,Williams2005}).  In order to relate the observed flux to the mass of the emitting material, a mass opacity $\kappa_{\nu}$ and a temperature $T_{\rm dust}$ have to be assumed  (cf. \citealt{Hildebrand1983}). To obtain the total (gas+dust) mass, the estimate of the dust mass has to be scaled up by the gas-to-dust mass ratio. Usually the interstellar medium (ISM) value of 100 is assumed, but it is unclear whether this is applicable to protoplanetary disks.

The gaseous component of the disk can be traced using carbon monoxide (CO), the second most abundant molecule in the disk, which is readily detected towards many protoplanetary disks (e.g., \citealt{Dutrey1996,Thi2001,Dent2005,Panic2008,WilliamsBest2014}). 
To derive a gas mass from CO observations, a relation has to be found between the amount of CO  in the disk and the amount of H$_2$. In the simplest case where most of the available carbon is contained in gas-phase CO, the fractional abundance of CO is $n( {\rm CO})/n({\rm H}_2) \sim 2\cdot10^{-4}$, in line with observations of warm dense clouds (e.g., \citealt{Lacy1994}).  

In protoplanetary disks there are two processes that decrease the CO abundance below its maximum value \citep{Zadelhoff2001}. In the upper layer of the disk, CO is being destroyed through photodissociation by ultraviolet (UV) photons. In the midplane, where temperatures are below $\sim 20$ K, CO freezes out onto the dust grains. As a result of these non-linear processes, thermochemical models have to be used to relate CO observations to the total gas mass. These processes are well understood and they are implemented in current thermochemical models. However, studies that include these effects find that overall gas phase carbon must still be additionally depleted by a factor 10-100. This leads to low CO emission even for massive disks (e.g., \citealt{Bruderer2012,Favre2013,Du2015,Kama2016,Bergin2016}).  

Several processes have been proposed to explain the underabundance of gas-phase carbon in regions where freeze-out is not important. A possible explanation \newtext{could be grain growth, where carbon has been locked up in large icy bodies that no longer participate in the gas-phase chemistry} (see, e.g., \citealt{Bergin2010,Bergin2016,Du2015,Kama2016}). Another cause could be the conversion of CO into more complex species. This process can happen in the gas-phase through reactions between CO and He$^+$, which extracts atomic carbon that can then be used to construct more complex molecules. These molecules have higher freeze-out temperatures than CO and freeze out onto the cold grains, thus lowering the apparent carbon abundance \citep{Aikawa1997,Favre2013,Bergin2014}. Similarly, ice chemistry can play an important role in converting CO into more complex organics, such as CH$_3$OH or turning it into CO$_2$ or CH$_4$ ice (see, e.g., Figure 3c in \citealt{Eistrup2016}).

Due to its high abundance, $^{12}$CO becomes optically thick at the disk surface, meaning it does not trace the bulk of the mass. Instead, less abundant CO isotopologues such as $^{13}$CO, C$^{18}$O and C$^{17}$O have to be used, which remain optically thin throughout the disk \citep{Zadelhoff2001,Dartois2003}. \cite{Visser2009} and \cite{Miotello2014} showed that C$^{18}$O/C$^{16}$O ratio is reduced beyond the $^{18}$O/$^{16}$O isotope ratio in disks due to isotope-selective photodissociation. Not taking this effect into account can change the estimated mass using C$^{18}$O up to one order of magnitude \citep{Miotello2014}.

Using the \textit{Herschel} Space Observatory, hydrogen deuteride (HD) has been detected towards TW Hya, DM Tau and GM Aur \citep{Bergin2013,McClure2016}. HD has been suggested as an alternative tracer of the disk gas mass. Because of their chemical similarities, the distribution of HD is expected to closely follow that of H$_2$. HD has a small dipole moment, so dipole transitions ($\Delta J = 1$) are allowed. The difference between the energy needed to excite the first rotational level of HD ($E/k_B \simeq128.5$ K) and the energy needed to excite the second rotational level of H$_2$ ($E/k_B \simeq549.2$ K), means that at a temperature of $T\sim20$ K the expected emission of HD is much larger than that of H$_2$.

The energetically lowest transition of HD, $J$=1-0, emits at 112 $\mu$m \citep{Muller2005}. This places the transition in a wavelength range where the dust continuum is bright, the atmospheric opacity is high and the production of low-noise detectors is expensive. All these reasons combined make detecting HD a challenge. Even with \textit{Herschel}, deep integrations were required to observe HD in disks. After the end of the \textit{Herschel} mission, there are currently no far-IR observatories capable of detecting HD in disks. However, this may change with the advent of future far-IR missions such as the proposed SPICA mission \citep{Nakagawa2014} and with the HIRMES instrument on SOFIA.

In this work, a simple HD chemistry is incorporated into the thermochemical code \underline{D}ust \underline{a}nd \underline{Li}nes (\texttt{DALI},\ \citealt{Bruderer2012,Bruderer2013}). 
In Section \ref{sec: model} a description of the code is given, including the deuterium chemistry that was implemented. A description of the models that were run is also given here. 
In Section \ref{sec: results} the dependence of the HD far-infrared emission on the disk gas mass is examined. In addition, the effect of other disk properties such as the vertical structure and the dust distribution on the HD emission is investigated. Furthermore, the results are examined in the light of possible future far-IR missions such as SPICA. 
The effect of prior knowledge of the disk vertical structure on the uncertainty of the HD based mass estimates is discussed in Section \ref{sec: discussion}. Additionally, the models are compared to the current observations. Of particular interest is the protoplanetary disk of TW Hya ($59.54\pm 1.45$ pc, \citealt{AstraatmadjaBailerJones2016}), where the gas mass estimated using HD is more than one order of magnitude higher than the gas mass estimated using C$^{18}$O \citep{Favre2013}. Different studies have shown that the TW Hya disk seems genuinely depleted in volatile carbon and oxygen (e.g.,  \citealt{Bergin2010,Bergin2013,Bergin2016,Hogerheijde2011,Favre2013,Du2015,Kama2016}). Recently, new observations of the CO isotopologues have become available (see, e.g., \citealt{Schwarz2016}). Therefore, the case of TW Hya is revisited using a combined HD and CO isotopologue analysis. The conclusions can be found in Section \ref{sec: conclusions}.


\section{Model}
\label{sec: model}
The thermochemical code \underline{D}ust \underline{a}nd \underline{Li}nes (\texttt{DALI})  was used to calculate the line fluxes for the disk models \citep{Bruderer2012,Bruderer2013}. \texttt{DALI} is a 2D physical-chemical code that for a given physical disk model calculates the thermal and chemical structure self-consistently. In addition to the disk density structure, a stellar spectrum has to be provided in order to determine the UV radiation field inside the disk. The computation proceeds through three steps. First the dust temperature structure and the internal radiation field (from UV- to mm-wavelengths) are determined by solving the continuum radiative transfer using a 2D Monte Carlo method. Next, the abundances of the molecular and atomic species at each point of the disk are obtained from the solution of the time-dependent chemistry. The atomic and molecular excitations are then computed using a non-LTE calculation. Given the excitation levels, the gas temperature is determined by balancing the heating and cooling processes. Due to the fact that both the chemistry and the excitations depend on temperature, the problem is solved iteratively until a self-consistent solution is found. Lastly, the model is ray-traced to construct spectral image cubes and line profiles. A more detailed description of the code can be found in Appendix A of \cite{Bruderer2012}. 

\subsection{Density structure}
\label{sec: density structure}
The density structure used in the code follows the simple parametric model proposed by \cite{Andrews2011}. This model is based on the assumption that the disk structure is determined by viscous accretion, where the viscosity goes as $\nu \propto R^{\gamma}$ \citep{LyndenBellPringle1974,Hartmann1998}.
The resulting surface density is given by 
\begin{equation}
\label{eq: surface density}
\Sigma_{\rm gas}(R) = \Sigma_c \left( \frac{R}{R_c}\right)^{\gamma} \exp \left[ -\left(\frac{R}{R_c}\right)^{2-\gamma}\right].
\end{equation}
Here $\Sigma_c$ is the surface density at the characteristic radius $R_c$.

\noindent The vertical structure is given by a Gaussian density distribution, as motivated by hydrostatic equilibrium \citep{KenyonHartmann1987} 
\begin{align}
\label{eq: vertical density}
n(R,z) &= \frac{1}{\sqrt{2 \pi}}\frac{1}{H} \exp \left[ -\frac{1}{2}\left(\frac{z}{H}\right)^2\right]\\
        &= \frac{1}{\sqrt{2 \pi}}\frac{1}{Rh} \exp \left[ -\frac{1}{2}\left(\frac{z/R}{h}\right)^2\right],
\end{align}
where $H = Rh$ is the physical height of the disk and the scale height angle $h$ is parametrized by
\begin{equation}
\label{eq: scale height}
h = h_c \left(\frac{R}{R_c}\right)^{\psi}.
\end{equation}
Here $h_c$ is the scale height at $R_c$ and $\psi$ is known as the flaring angle. \newtext{Equation \eqref{eq: scale height} is a representation for a physical disk, where the vertical structure is set by balancing gravity and the vertical pressure gradient at each point in the disk.}

\subsection{Dust settling}
\label{sec: dust settling}

In order to include dust settling into the model, the dust grains are split into two populations:
\begin{itemize}
    \item \textit{small grains} (0.005-1 $\mu$m) are included with a (mass) fractional abundance $1-f_{\rm large}$. These are located throughout the vertical extent of the disk.
    \item \textit{large grains} (1-$10^3\ \mu$m) are included with a fractional abundance $f_{\rm large}$. They are limited to a vertical region with scale height $\chi h;\ \chi < 1$, simulating the effect of dust settling.
\end{itemize}

\begin{figure}[htb]
\centering
\includegraphics[trim={0cm 3.5cm 0cm 2cm},clip,width = \columnwidth]{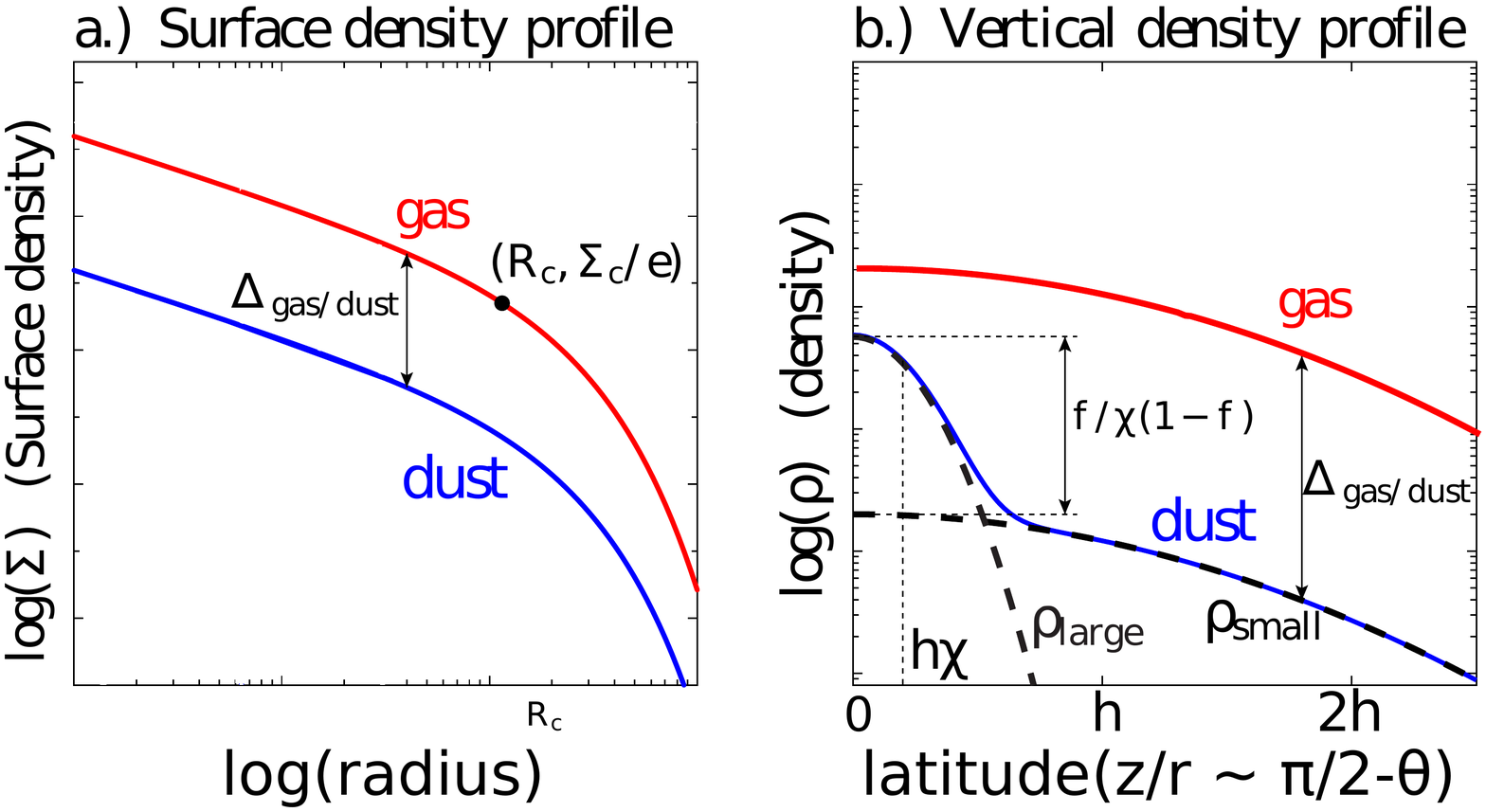}
\caption{\label{fig: dali model structure}Gas and dust density structure assumed in \texttt{DALI}}
\end{figure}
\noindent Note that in the current version of \texttt{DALI} the gas surface density profile is only coupled to the density profile of the small grains, in contrast to earlier versions of \texttt{DALI}. The resulting disk structure is summarized in Figure \ref{fig: dali model structure}.

A standard ISM dust composition is assumed for the dust opacities following \cite{WeingartnerDraine2001}, in line with \cite{Bruderer2013} (see e.g. his Figure 2). The mass opacities for absorption at 112 $\mu$m and 56 $\mu$m for both dust grain populations are given in Table \ref{tab: opacities}. 

\begin{table}[htb]
  \centering   
  \caption{\label{tab: opacities}Dust mass opacities}
  \begin{tabular*}{0.85\columnwidth}{lcc}
    \hline\hline
    & wavelength [$\mu$m] & $\kappa_{\rm abs.}$ [cm$^2$ g$^{-1}$]\\
    \hline
    small grains & 112  & 29.9  \\ 
                 & 56   & 154 \\  
    large grains & 112 & 30.0  \\
                 & 56  & 46.3\\ 
    \hline
  \end{tabular*}
\end{table}

\subsection{Chemical network}
\label{sec: chemical network}

For the models in this work the list of chemical species adopted in \cite{Miotello2014} was extended to include HD, D, HD$^+$ and D$^+$ as separate species. By adding the element D to H, He, $^{12}$C, $^{13}$C, N, $^{16}$O, $^{17}$O, $^{18}$O, Mg, Si, S and Fe, the total number of species was increased to 280. The reaction types included are H$_2$ and HD formation on dust, freeze-out, thermal and non-thermal desorption, hydrogenation of simple species on ices, gas-phase reactions, photodissociation, X-ray and cosmic-ray induced reactions, PAH/small grain charge exchange/hydrogenation, and reactions with H$_2^*$ (vibrationally excited H$_2$). The details of the implementation of these reactions are described in Appendix A.3.1 of \cite{Bruderer2012}. 

To properly simulate HD integrated line fluxes, simple deuterium chemistry has been added to the chemical network. The only D-bearing species considered for the chemical calculation are D, D$^+$, HD, and HD$^+$. Accordingly, the important reactions regulating their abundances have been included, with rate coefficients from \cite{RobertsMillar2000,Walmsley2004}. Photodissociation of HD was included following \cite{GloverJappsen2007}. The photodissociation rate for a radiation field with a flat spectrum is given by Eq. 51 in \cite{GloverJappsen2007}:
\begin{equation}
\label{eq: HD photodissociation}
R_{\rm diss,HD} = 1.5\times10^{-9} I(\nu)\ \mathrm{s}^{-1}.
\end{equation}
Here $I(\nu)$ is the mean intensity of the radiation field. Note that the prefactor has units such that the photodissociation rate has units [s$^{-1}$].

HD self-shielding was also included in the model using the shielding function of H$_2$ but slightly shifted in wavelength. Finally HD formation onto grains and ion-exchange reactions with metals and PAHs have been included, with the same rate coefficients as that assumed for the analogue reactions with H instead of D. The reactions involving D-bearing species added in the chemical network are presented in Table \ref{tab: deuterium chemistry} in Appendix \ref{app: deuterium table}.

\subsection{Grid of models}
\label{sec: our models}
As a basis for the models, the physical structure of the TW Hya disk from the \cite{Kama2016} model is used. A combination of spatially and spectrally resolved and unresolved line fluxes, in combination with observations of the spectral energy distribution (SED), was used to constrain the parameters of their model, which can be found in Table \ref{tab: grid parameters}. Note that \cite{Kama2016} find a strongly flared vertical structure for TW Hya. Also, their model has a radially steeper temperature profile and smaller outer radius than the models of \cite{Cleeves2015}. 

In total, over a 160 models were run using \texttt{DALI}. The employed chemical network is the expanded version of that used by \cite{Miotello2014}, where simple deuterium chemistry has been considered (see Section \ref{sec: chemical network}). The input total disk masses are M$_{\mathrm{disk}} = [2.6\cdot10^{-5},\ 2.6\cdot10^{-4},\ 7.7\cdot10^{-4},\ 2.6\cdot10^{-3},\ 7.7\cdot10^{-3},\ 2.3\cdot10^{-2},\ 7.7\cdot10^{-2},\ 2.6\cdot10^{-1}]$ M$_{\odot}$. For each mass, 20 models with different vertical structures were ran: $h_c \in [0.05, 0.1, 0.2, 0.3], \psi \in[0.1, 0.2, 0.3, 0.4, 0.5]$ (cf. equation \ref{eq: scale height}). The other parameters have been kept fixed to the values assumed by \cite{Kama2016} (Table \ref{tab: grid parameters}). For the overall carbon and oxygen abundances, typical ISM values of [C]/[H]$_{\rm ISM} = 1.35\cdot10^{-4}$, [O]/[H]$_{\rm ISM} = 2.88\cdot10^{-4}$ are used \citep{Bruderer2012}.

\newtext{The stellar spectrum for TW Hya from \cite{Cleeves2015} was used as the central source of radiation. It closely resembles a 4000 K blackbody with a 10000 K blackbody component to represent the observed UV excess (following the prescription in \citealt{Kama2016a},  see also Figure 3 and Table 3 in \citealt{Kama2016}).}

The time-dependent chemistry is run for 1 Myr, which is less than the expected age for TW Hya, but the HD emission is not sensitive to the chemical age.

The overall deuterium abundance [D]/[H] of the gas in the disk has a direct impact on the HD fluxes. \cite{Prodanovic2010} used observations of the atomic D and H lines in the diffuse ISM to find $(\mathrm{[D]/[H]})_{\rm ISM} \geq (2.0 \pm 0.1)\times10^{-5}$ for the local ISM (i.e., within $\sim 1-2$ kpc of the Sun). The uncertainties due to the error on the assumed [D]/[H] are much smaller than the spread in HD fluxes due to the disk structure. Note that for older star forming regions such as Orion the deuterium abundance is lower, due to the fact that a fraction of the deuterium has been burned up in the cores of stars \citep{Wright1999,Howat2002}.

\begin{table}[htb]
  \centering   
  \caption{\label{tab: grid parameters}\texttt{DALI} parameters of the physical model.}
  \begin{tabular*}{0.95\columnwidth}{ll}
    \hline\hline
    Parameter & Range\\
    \hline
     \textit{Chemistry}&\\
     Chemical age & 1 Myr\\
     $\mathrm{[D]/[H]}$ & $2\cdot10^{-5}$ \\
     \textit{Physical structure} &\\ 
     $\gamma$ &  1.0\\ 
     $\psi$ & [0.1,0.2,\textbf{0.3},0.4,0.5]\\ 
     $h_c$ &  [0.05,\textbf{0.1},0.2,0.3] rad\\ 
     $R_c$ & 35 AU \\
     $R_{\mathrm{out}}$ & 200 AU\\ 
     M$_{\mathrm{disk}}$ &$2.6\cdot10^{-5},\ 2.6\cdot10^{-4},\ 7.7\cdot10^{-4},$\\ 
     & $2.6\cdot10^{-3},\ 7.7\cdot10^{-3},\ \mathbf{2.3\cdot10^{-2}}$ \\
     & $7.7\cdot10^{-2},\ 2.6\cdot10^{-1}$ M$_{\odot}$\\
     \textit{Dust properties} &\\
     Gas-to-dust ratio & 200 \\
     $f_{\mathrm{large}}$ & 0.99 \\
     $\chi$ & 0.2 \\
     $f_{\mathrm{PAH}}$ & 0.001 \\
     \textit{Stellar spectrum$^1$} & \\
     type & T Tauri \\
     L$_{\rm X}$ & 10$^{30}$ erg s$^{-1}$ \\
     \newtext{L}$_{\rm FUV}$ & $2.7\cdot10^{31}$ \newtext{erg s}$^{-1}$ \\
     L$_{*}$ & 0.28 L$_{\odot}$ \\
     $\zeta_{\rm cr}$ & $10^{-19}\ \mathrm{s}^{-1}$\\
     \textit{Observational geometry}&\\
     $i$ & 6$^{\circ}$\\
     d & 140 pc\\
    \hline
  \end{tabular*}
  \captionsetup{width=.9\columnwidth}
  \caption*{\footnotesize{Bold numbers denote the best fit values for TW Hya from \cite{Kama2016}. The deuterium fraction was taken from \cite{Prodanovic2010}.$^1$ \newtext{Spectrum of TW Hya} \citep{Cleeves2015}. }}
\end{table}


\section{Results}
\label{sec: results}

\subsection{HD flux vs. disk gas mass}
\label{sec: HD flux vs. disk gas mass}

\begin{figure*}[htb]
\centering
\includegraphics[height=0.3\textheight]{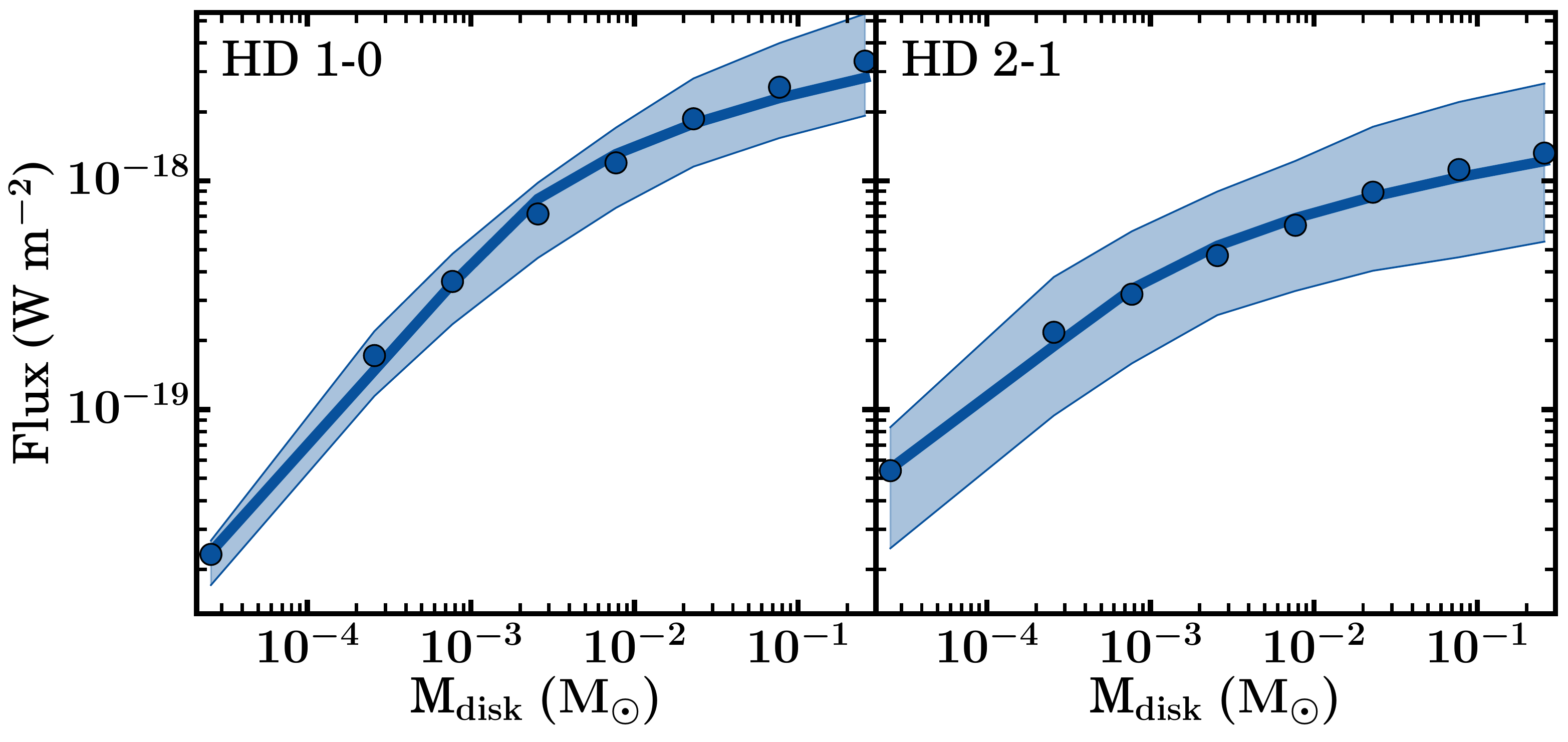}
\caption{\label{fig: fluxMass}\textbf{Left panel:} Integrated line flux (calculated at 140 pc) of the HD 1-0 transition as a function of disk mass. The points show the median flux of the models in that mass bin. The blue shaded region shows fluxes between the 16$^{\rm th}$  and 84$^{\rm th}$ percentiles in the same mass bin. The blue line shows the fit of equation \eqref{eq: model flux-mass} using the values from Table \ref{tab: fit coefficients}. 
\textbf{Right panel:} Integrated line flux (calculated at 140 pc) of the HD 2-1 transition as function of disk mass.}
\end{figure*}

The first step in investigating the viability of HD as a tracer of the disk gas mass is examining the behaviour of its emission under variation of the gas mass. This is presented in Figure \ref{fig: fluxMass}, where the integrated HD 1-0 and 2-1 line fluxes are plotted as function of disk mass. In the left panel of Figure \ref{fig: fluxMass} the blue dots shows the median HD 1-0 flux for each disk mass, while the coloured region presents fluxes between the 16$^{\rm th}$  and 84$^{\rm th}$ percentiles in each mass bin. The HD 1-0 line flux increases with mass, following a linear dependence in log-log space for the low-mass disks (M$_{\rm disk} \leq 10^{-3}\ \mathrm{M}_{\odot}$). 

Towards the high mass regime, the relation flattens. There are two reasons for this change in behaviour. For disk masses above 0.001 M$_{\odot}$ the HD 1-0 line emission starts to become optically thick, which means that the HD emission no longer traces the full gas reservoir. In addition, high mass disks have larger densities, which decreases the gas temperature in the disk. \cite{Miotello2016} found a similar decrease in line luminosity for $^{13}$CO and C$^{18}$O for high mass disks, which they attributed to optical depth for $^{13}$CO and increased freeze-out for C$^{18}$O.

The right panel of Figure \ref{fig: fluxMass} shows that the flux-mass relation for the HD 2-1 line is flatter compared to HD 1-0. This is likely due to increased temperature dependence of the HD 2-1 line, compared to the HD 1-0 line. The increased temperature dependence is reflected in the increased sensitivity of HD 2-1 to  the vertical structure of the disk, as indicated by the large flux variations in each mass bin. 

The behaviour of the flux-mass relation can be expressed with an analytical relation, such as:
\begin{equation}
\label{eq: model flux-mass}
        F_{\rm HD}  = \begin{cases}
                            A\left( \mathrm{M}_{\rm disk}/\mathrm{M}_{\rm tr}\right)^\alpha,& \text{if } \mathrm{M}_{\rm disk} \leq \mathrm{M}_{\rm tr}\\
                            A +  B\log_{10}\left(\mathrm{M}_{\rm disk}/\mathrm{M}_{\rm tr}\right), & \text{otherwise}
    \end{cases}
\end{equation}

\noindent where M$_{\rm tr}$ is defined as the mass where the flux-mass relation starts to flatten.

This model was fitted to the median fluxes of both HD lines. The resulting coefficients can be found in Table \ref{tab: fit coefficients}. Note that the powerlaw slopes for both HD lines are sublinear. This is likely a result from the fact that the HD emission is produced by warm gas only. 

\begin{table}[htb]
  \centering   
  \caption{\label{tab: fit coefficients}Fit coefficients of the HD flux-mass relation.}
  \begin{tabular*}{0.75\columnwidth}{lcc}
    \hline\hline
    & HD 1-0 & HD 2-1\\
    \hline
    $\alpha$ & 0.79 & 0.54 \\
    A & 5.44$\cdot10^{-19}$ & 2.83$\cdot10^{-19}$ \\
    B & 9.95$\cdot10^{-19}$ & 3.50$\cdot10^{-19}$ \\
    M$_{\rm tr}$ (M$_{\odot}$) & 1.32$\cdot10^{-3\ }$ & 5.43$\cdot10^{-4\ }$ \\
    \hline
  \end{tabular*}
  \captionsetup{width=.7\columnwidth}
  \caption*{\footnotesize{Coefficients fitted to HD fluxes calculated at 140 pc. For a different distance $d$, the coefficients $A$ and $B$ scale as $A' = (140/d)^2\cdot A$. }}
\end{table}

\subsection{HD emitting layers}
\label{sec: HD emission maps}

\begin{figure}
\centering
\begin{subfigure}{\columnwidth}
\includegraphics[width=\columnwidth]{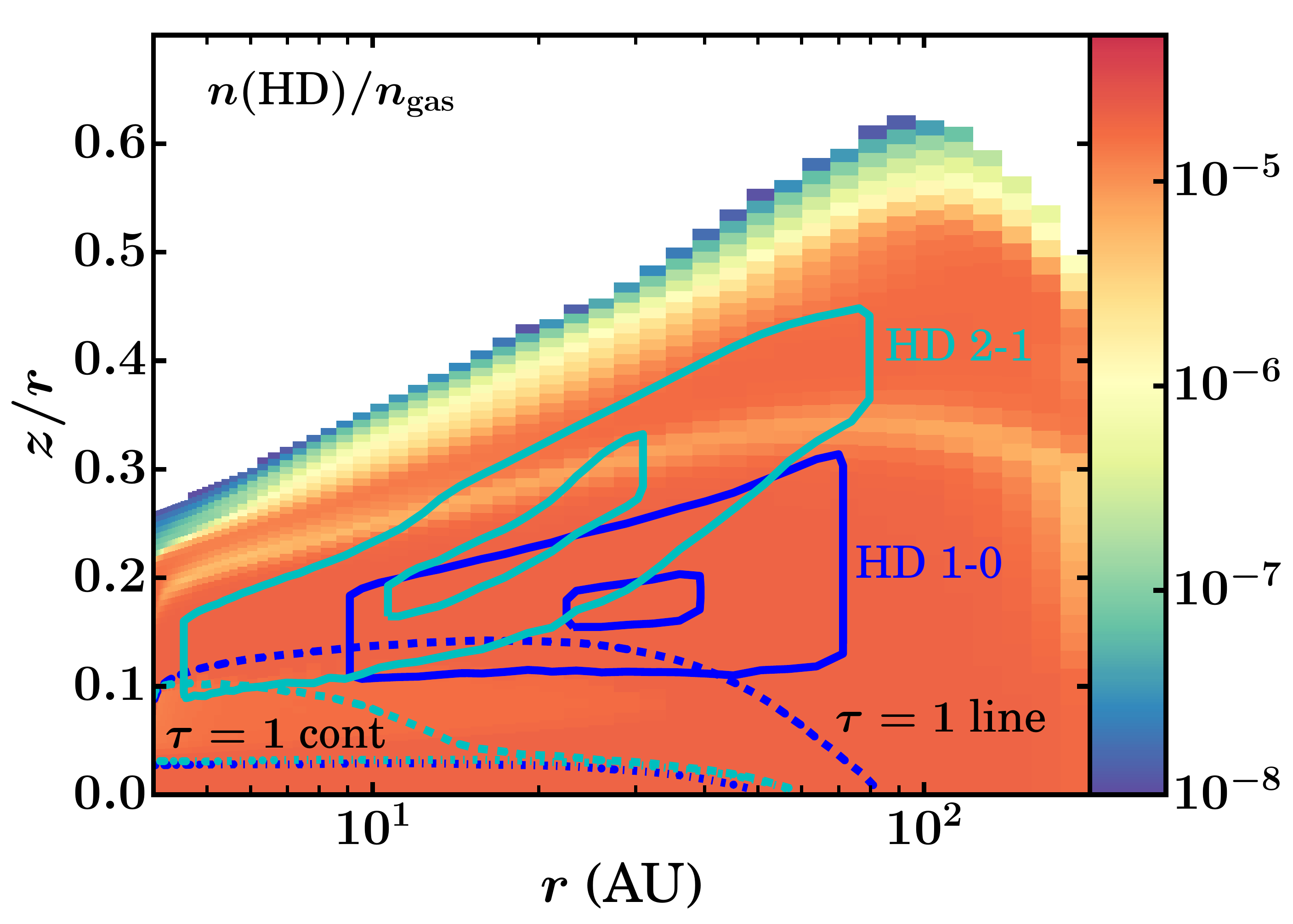}
\end{subfigure}
\caption{\label{fig: HDdistribution}HD abundance structure for a disk model with M$_{\rm disk} = 2.3\cdot10^{-2}\ \mathrm{M}_{\odot}$, $\psi = 0.3$ and $h_c = 0.1$. Colour indicates the number density of HD with respect to the total gas density. The coloured lines correspond the HD 1-0 (blue) and HD 2-1 (light blue) respectively. Solid contours indicate where 25\% and 75\% of the emission originates from. The dashed (dashed-dotted) lines show the $\tau=1$ surface for the line and continuum opacity, respectively.}
\end{figure}

The 2D abundance of HD in a typical model is presented in Figure \ref{fig: HDdistribution}. This distribution is the outcome of the deuterium chemical calculation discussed in Section \ref{sec: chemical network}. Note that $n(\mathrm{HD})/n_{\rm gas} = 2\cdot10^{-5}$ throughout most of the disk, indicating that all of the available deuterium is locked up in HD. The exceptions to this rule are in the upper layer of the disk, where HD is photodissociated and in an intermediate layer where HD abundance is decreased by a factor of $\sim$2. This layer is due to a combination of chemical processes. It should be noted here that this layer is very thin and only covers a very small fraction of the HD emission area shown in Figure \ref{fig: HDdistribution}. Therefore the depletion in this layer has no relevant effect on the HD flux.

The solid blue lines in Figure \ref{fig: HDdistribution} denote where 25\% and 75\% of the HD 1-0 emission is produced. This region is elevated above the midplane ($0.1 \leq z/r \leq 0.25$), indicating that most of the HD 1-0 emission is produced by warm gas, with an average temperature of $30-70$ K (cf. Figure \ref{fig: Temperatures}). The emitting region extends radially from $r_{\rm in} \approx 9$ AU up to $r_{\rm out}\approx 70$ AU for M$_{\rm disk} = 2.3\cdot10^{-2}\ \mathrm{M}_{\odot}$. The inner and outer radii vary slightly vertical structure (cf. Figure \ref{fig: emission vertical structure}). Figure \ref{fig: emission mass} shows that when the disk gas mass is increased the emitting region shifts outward, from $(r_{\rm in}, r_{\rm out}) \approx (1\ \mathrm{AU},40\ \mathrm{AU})$ for M$_{\rm disk} = 2.3\cdot10^{-5}\ \mathrm{M}_{\odot}$ to $(r_{\rm in}, r_{\rm out}) \approx (10\ \mathrm{AU},100\ \mathrm{AU})$ for M$_{\rm disk} = 2.3\cdot10^{-1}\ \mathrm{M}_{\odot}$. This is a result of the optically thick HD zone extending to larger radii as the gas mass increases.

The $\tau=1$ surfaces for the line and continuum optical depth are shown in dashed and dashed-dotted lines, respectively. They show that continuum optical depth has no effect on the line emission. This is likely due to the fact that most of the dust mass is in large grains, which are settled to the midplane in contrast with the models by \cite{Bergin2014}. The dashed line shows that for this model with M$_{\rm disk} = 2.3\cdot10^{-2}$ M$_{\odot}$, the HD emission starts to become optically thick, which is in line with the flattening of the slope seen in Figure \ref{fig: fluxMass}.

For the HD 2-1 emission, shown in light blue in Figure \ref{fig: HDdistribution}, the emission originates even higher up in the disk, where gas temperatures are in the range of $70-100$ K (cf. Figure \ref{fig: Temperatures}). The contour for the line optical depth show that the HD 2-1 emission starts to become optically thick in the inner part of the disk for M$_{\rm disk} = 2.3\cdot10^{-2}$ M$_{\odot}$.

\subsection{Influence of the vertical structure}
\label{sec: influence of the vertical structure}

\begin{figure*}[tbh]
\centering
\includegraphics[height = 0.35\textheight]{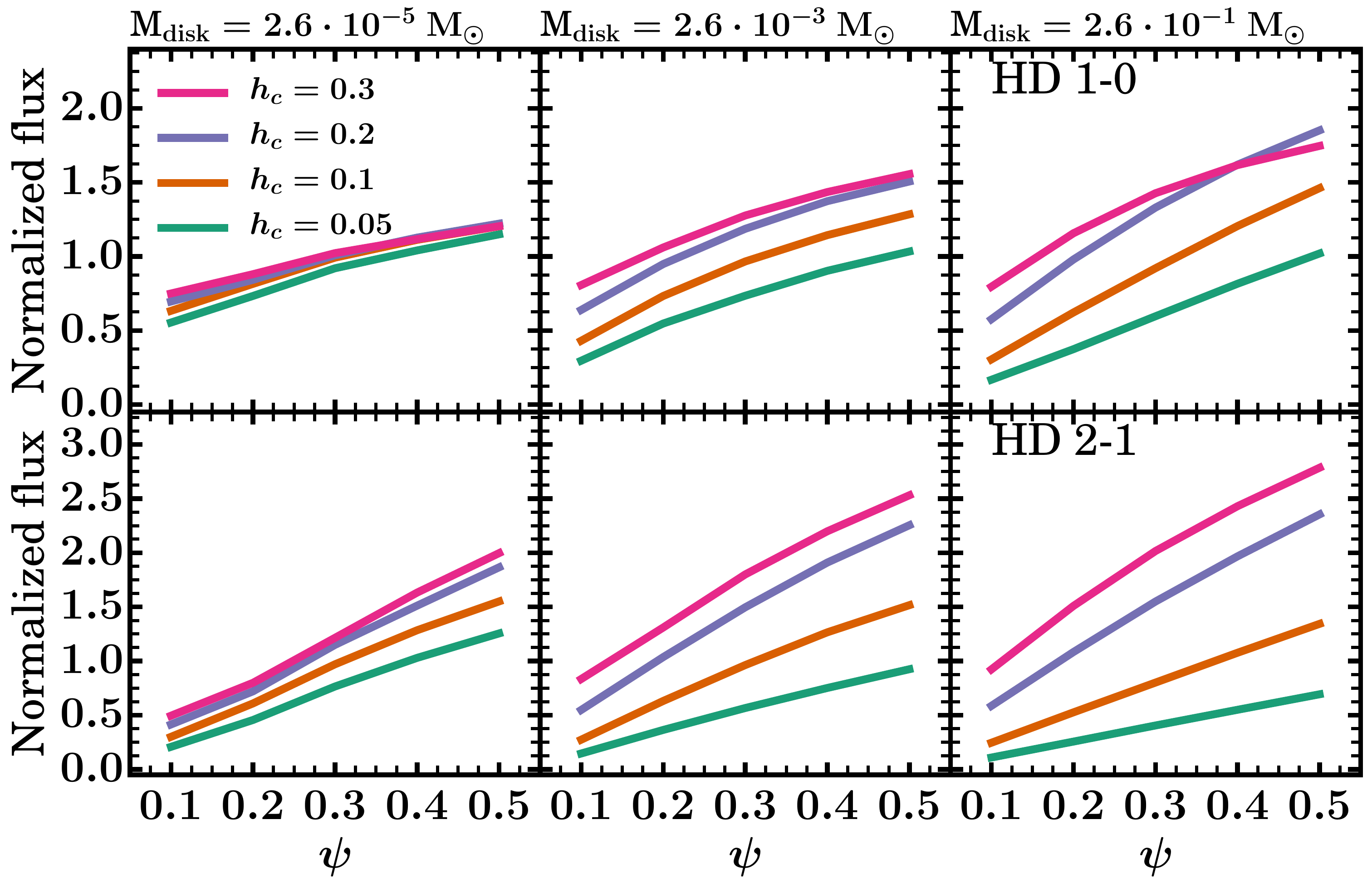}
\caption{\label{fig: fluxVS} Integrated HD 1-0 (top) and HD 2-1 (bottom) line fluxes as function of flaring angle $\psi$ for different scale heights $h_c$. The fluxes are normalized with respect to median flux of the models with the same disk mass (cf. Figure \ref{fig: fluxMass}).}
\end{figure*}

The strength of the HD emission depends on the amount of material and the temperature of the material. In protoplanetary disks the temperature structure is set by the vertical structure of the disk \citep{ChiangGoldreich1997}. The effect of the vertical structure on the HD 1-0 and 2-1 fluxes is presented in Figure \ref{fig: fluxVS}. It shows the HD fluxes as function of the flaring angle $\psi$ for different scale heights $h_c$ (colours). This was done for three different mass bins ($2.3\cdot10^{-5}\ \mathrm{M}_{\odot}$, $2.3\cdot10^{-3}\ \mathrm{M}_{\odot}$ and $2.3\cdot10^{-1}\ \mathrm{M}_{\odot}$). The fluxes in each mass bin are normalized with respect to the median flux in the same bin (cf. the blue dots in Figure \ref{fig: fluxMass}). 

The normalized HD 1-0 integrated line fluxes are presented in top panels of Figure \ref{fig: fluxVS}. It shows that the HD 1-0 flux increases with increasing $\psi$. The relation is roughly linear, with a slope that is independent of $h_c$ and M$_{\rm disk}$.

The spacing between the different $h_c$ lines indicate that the HD 1-0 flux also increases systematically with $h_c$. The shape of the flux-$h_c$ relation depends only weakly on $\psi$. It does depend strongly on disk mass, with the influence of $h_c$ on the flux increasing as function of disk mass.

For HD 2-1, presented in the bottom panels of Figure \ref{fig: fluxVS}, the dependencies of the flux on $h_c$ and $\psi$ are stronger than those found for HD 1-0. The bulk of the HD 2-1 emission is produced by gas at temperatures of $70-100$ K (cf. Figure \ref{fig: Temperatures}). This gas is located higher up in the disk. As the optically thin HD 2-1 emission depends on both the temperature and the density, the location of the HD 2-1 emitting material varies strongly with vertical structure, as can be seen in Figure \ref{fig: emission vertical structure}. This is in contrast to the HD 1-0 emitting material, which stays at roughly the same location, independent of the vertical structure.

Figure \ref{fig: fluxVS} shows that the maximum variation in flux due to the different vertical structure increases with disk mass, from $0.75\ \times$ the median flux for $\mathrm{M}_{\rm disk} \sim 10^{-5}\ \mathrm{M}_{\odot}$ up to $1.9\ \times$ the median flux for $\mathrm{M}_{\rm disk} \sim 10^{-1}\ \mathrm{M}_{\odot}$. For the HD 2-1 integrated line flux, shown in the bottom panel of Figure \ref{fig: fluxVS}, the maximum variation in flux is both larger than for HD 1-0 and is less dependent on disk mass. The maximum variation of the flux is $1.9\ \times$ the median flux for $\mathrm{M}_{\rm disk} \sim 10^{-5}\ \mathrm{M}_{\odot}$, increasing to $2.8\ \times$ the median flux for $\mathrm{M}_{\rm disk} \sim 10^{-1}\ \mathrm{M}_{\odot}$. As mentioned in Section \ref{sec: HD emission maps}, the regions where the HD lines are optically thick have larger outer radii as the gas mass is increased. On the one hand this pushes the HD emitting regions upwards to parts of the disk that are more sensitive to the vertical structure. On the other hand, the increased size of the optically thick HD zone means that, for massive disks, a larger mass fraction of HD will be much warmer as the vertical structure changes. \newtext{The uncertainty in derived gas masses from HD due to vertical structure is further quantified in Section \ref{sec: determining the disk gas mass}}.

\newtext{In the above models, the vertical structure is parametrized according to Equation \eqref{eq: scale height}. Another approach would be to calculate the vertical structure using hydrostatic equilibrium from the computed gas temperature structure. The effect of this approach on the HD fluxes is examined in Appendix \ref{app: effects of including hydrostatic equilibrium} and is found to be within the flux variations shown in Figure \ref{fig: fluxVS}.}

\subsection{Influence of the large grains}
\label{sec: influence of the large grains}

\begin{figure}[tbh]
\centering
\includegraphics[width =0.9\columnwidth]{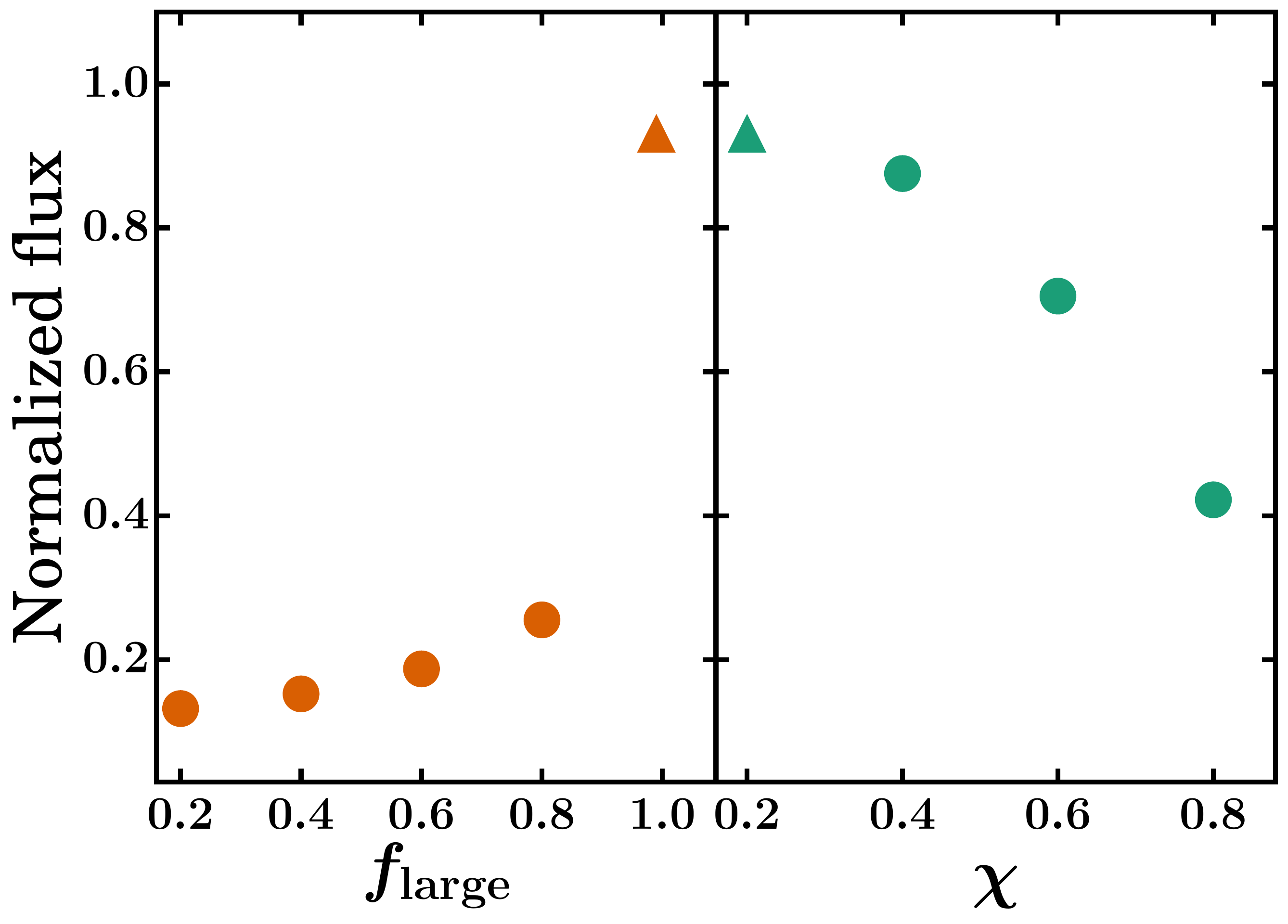}
\caption{\label{fig: parameters} 
Integrated line flux of HD 1-0 as function of dust parameters for M$_{\rm disk} = 2.3\cdot10^{-2}\ \mathrm{M}_{\odot}$. Models with different $f_{\rm large}$ and $\chi$ are shown in orange and green, respectively. The fluxes are normalized to the median flux of the $\mathrm{M}_{\rm disk} = 2.3\cdot10^{-2}\ \mathrm{M}_{\odot}$ mass bin. The triangular marker denotes the value of the parameter ($f_{\rm large}$ or $\chi$) in the fiducial model of the TW Hya disk (\citealt{Kama2016}, cf. Table \ref{tab: grid parameters}).    
}
\end{figure}

For the models discussed in Section \ref{sec: our models} only three parameters (M$_{\rm disk}$, $\psi$ and $h_c$) were varied. However, several other parameters may also affect the HD flux. Their role is investigated here. In particular, the large dust grains in the disk can affect the HD flux, both by affecting heating and by changing the optical depth of the dust at the wavelengths where HD emits. Two dust parameters are examined here: the mass fraction of the large grains $f_{\rm large}$ and the dust settling parameter of the large grains $\chi$ (cf. Section \ref{sec: dust settling}).  In both cases, the mass and vertical structure of the disk were fixed to the values from \cite{Kama2016}. The resulting HD 1-0 fluxes, normalized to the median flux of the M$_{\rm disk} = 2.3\cdot10^{-2}\ \mathrm{M}_{\odot}$ mass bin, are shown in Figure \ref{fig: parameters}.

Figure \ref{fig: parameters} shows that the HD 1-0 flux increases with dust settling (i.e., when the large grains are more settled towards the midplane). This can be understood by the fact the dust optical depth at a fixed height in the disk increases when the large grain population (cf. Section \ref{sec: dust settling}) is less settled. This is especially the case in the model for TW Hya, where most of the dust mass is in these grains. 

A similarly large effect is found when the mass fraction of the large grains is changed. When $f_{\rm large}$ decreases, the amount of small grains increases significantly, which increases the opacity at 112 $\mu$m. 

Comparing these results with those from the previous section, it is clear that both the large grains and the vertical structure have similar effects on the HD 1-0 line flux. It should be noted however that both $\chi$ and $f_{\rm large}$ are constrained by the spectral energy distribution (e.g., \citealt{DAlessio2001,Chiang2001}), so their values cannot be chosen arbitrarily.

\subsection{Influence of the gas-to-dust ratio}
\label{sec: influence of the gas-to-dust ratio}

\begin{figure}[tbh]
\centering
\includegraphics[width =0.9\columnwidth]{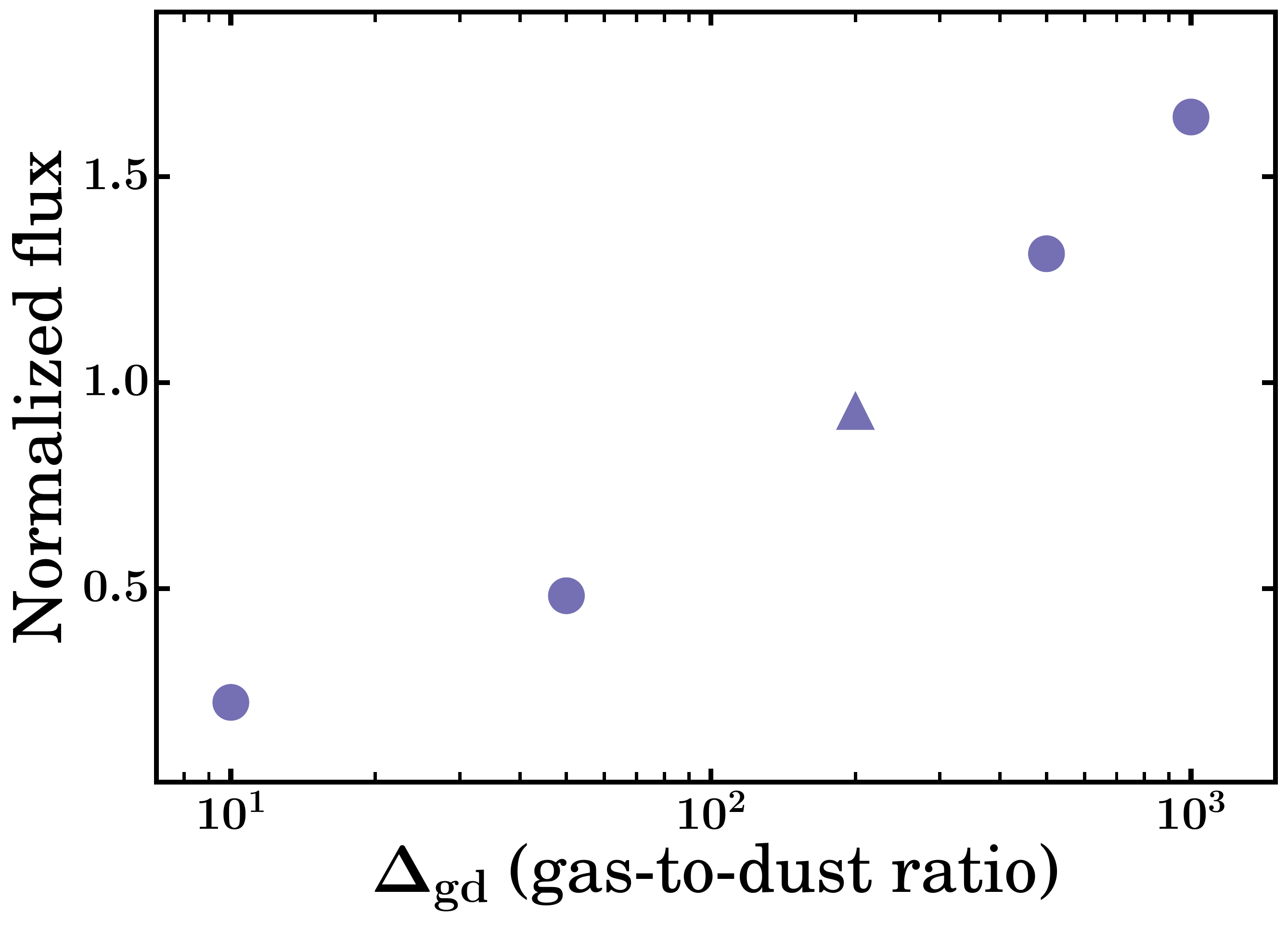}
\caption{\label{fig: gdratio} 
\newtext{Integrated line flux of HD 1-0 as function of gas-to-dust mass ratio for fixed gas mass( M$_{\rm gas} = 2.3\cdot10^{-2}\ \mathrm{M}_{\odot}$). The fluxes are normalized to the median flux of the $\mathrm{M}_{\rm disk} = 2.3\cdot10^{-2}\ \mathrm{M}_{\odot}$ mass bin. The triangular marker denotes the value of $\Delta_{\rm gd}$ in the fiducial model of the TW Hya disk (\citealt{Kama2016}, cf. Table \ref{tab: grid parameters}). 
}}
\end{figure}

\newtext{Deriving accurate gas-to-dust mass ratios has been one of the motivations for measuring the gas mass directly (e.g., \citealt{Ansdell2016,Miotello2017}). However, the gas-to-dust ratio could also affect the HD emission directly, e.g., by changing the coupling between the dust and the gas. If the effect is strong it may prevent accurate measurements of the gas-to-dust ratio using gas masses derived from HD emission. For fixed gas mass of M$_{\rm gas} = 2.3\cdot10^{-2}\ \mathrm{M}_{\odot}$, Figure \ref{fig: gdratio} presents the HD 1-0 fluxes as functions of increasing gas-to-dust ratio (and therefore decreasing dust mass). The fluxes are normalized with respect to the median flux of the M$_{\rm disk} = 2.3\cdot10^{-2}\ \mathrm{M}_{\odot}$ mass bin. }

\newtext{The figure shows that the HD 1-0 flux increases systematically with $\Delta_{\rm gd}$. The primary reason for this is the decreased coupling between the gas and the dust for higher gas-to-dust ratios. As the gas in the HD emitting region is less efficient in cooling than the dust, this leads to slightly higher gas temperatures. Additionally, the increased dust mass for the models with a lower $\Delta_{\rm gd}$ enhances the dust opacities at the wavelengths where HD emits, which lowers the amount of observed line flux. However, inspection of the $\tau_{\rm dust} =1$ surface at 112 $\mu$m shows that the increase in opacity is relatively small.}   

\subsubsection{Determining gas-to-dust ratios using HD 1-0}
\label{sec: determining gas-to-dust ratios}

\begin{figure}
    \includegraphics[width=\columnwidth]{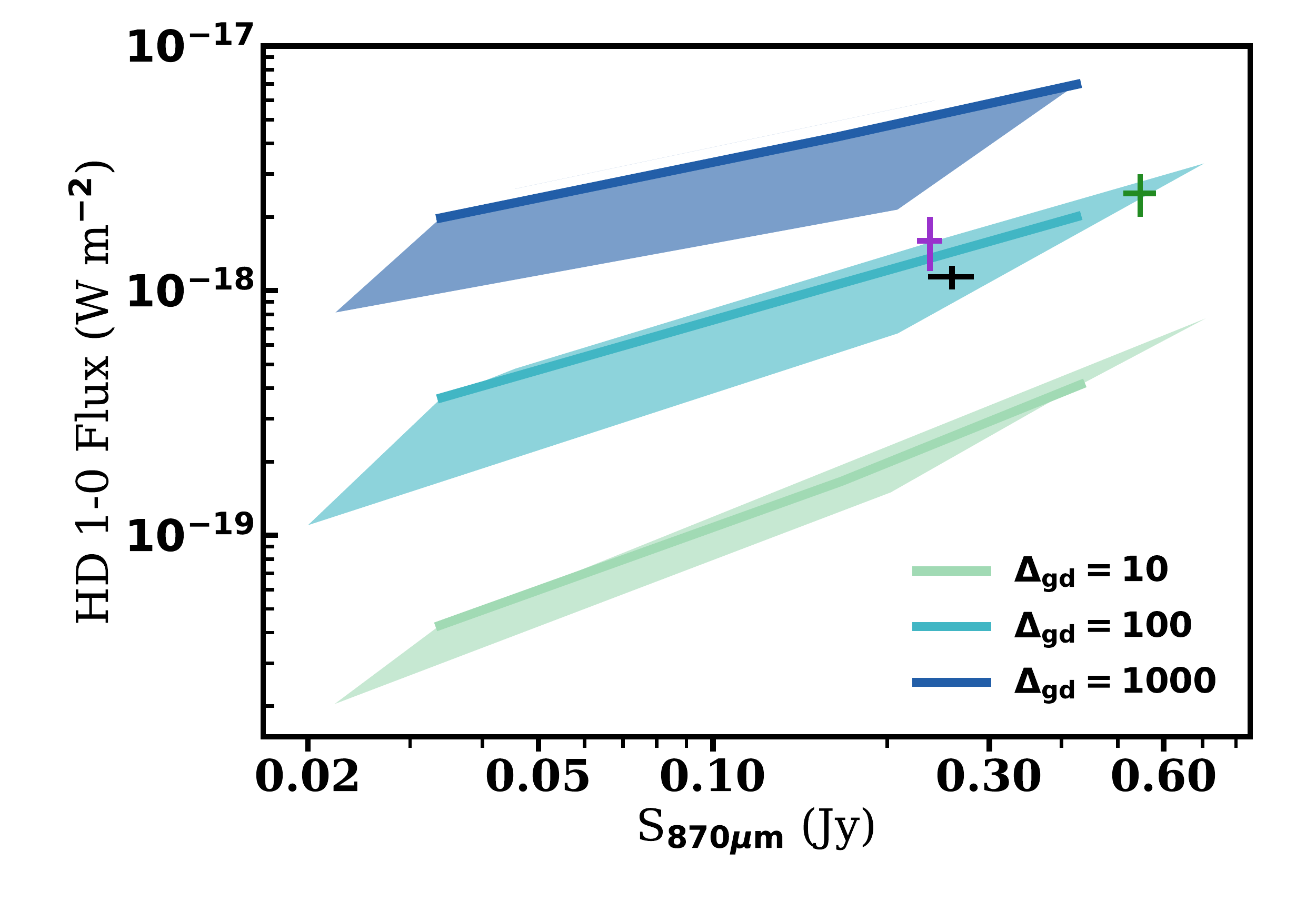}
    \caption{\label{fig: dust masses} \newtext{ HD 1-0 integrated fluxes and 870 $\mu$m continuum flux densities for the models described in Table \ref{tab: dust masses}. Colours indicate different gas-to-dust ratios, where the shaded region shows the flux variations due to different vertical structures. For each $\Delta_{\rm gd}$, the coloured line shows models with ($\psi,h_c$) = (0.3,0.1). The crosses denote observations for TW Hya (black; \citealt{Bergin2013,Andrews2012}), DM Tau (purple; \citealt{McClure2016, Andrews2013}) and GM Aur (green; \citealt{McClure2016,Andrews2013}). Note that both the model fluxes and the observations were scaled to a distance of 140 pc. }}
\end{figure}

\newtext{Figure \ref{fig: gdratio} shows that varying the gas-to-dust ratio by two orders of magnitude for a fixed gas mass only leads to a variation in the HD 1-0 line flux of $\sim 1.5 \times$ the median flux. In this analysis the dust mass is assumed to be a free parameter. However, in practice the dust mass of the observed disks will be constrained, e.g., by observation of their millimeter-continuum fluxes. A comparison of HD 1-0 fluxes and millimeter continuum fluxes for models with different $\Delta_{\rm gd}$ is shown in Figure \ref{fig: dust masses}. }

\newtext{Here the HD emission is combined with the millimeter continuum emission of the dust to provide a method for estimating the gas-to-dust ratio from observations. For this purpose an additional set of 27 models was run, where the dust mass, gas-to-dust ratio and the vertical structure were varied (cf. Table \ref{tab: dust masses}). Figure \ref{fig: dust masses} shows the HD 1-0 line flux, which traces the gas mass, plotted against the 870 $\mu$m continuum flux, which traces the dust mass. The lack of overlap between models with different gas-to-dust ratios indicates that combined observations of HD and the dust continuum can be used to determine $\Delta_{\rm gd}$. For the three disks for which HD 1-0 observations are available (i.e., TW Hya, DM Tau and GM Aur) gas-to-dust ratios are found to be $\Delta_{\rm gd} \sim 100$, close to the ISM value. This would suggest that indeed the lower values found from similar measurements made using $^{13}$CO are due to an underabundance of volatile carbon (see \citealt{Miotello2017} and references therein).}

\begin{table}[htb]
  \centering
  \caption{\label{tab: dust masses}Disk parameters for the fixed dust mass models.}
  \begin{tabular*}{0.95\columnwidth}{ll}
    \hline\hline
    Parameter & Range\\
    \hline
    \textit{Dust masses}\\
    M$_{\mathrm{dust}}$ &$10^{-5},\ 10^{-4},\ 10^{-3}$ M$_{\odot}$\\
    \textit{Gas-to-dust ratios}\\
    $\Delta_{\rm gd}$ & 10, 100, 1000 \\
    \textit{Gas masses$^1$}\\
    M$_{\mathrm{gas}}$ & $10^{-5} - 1$ M$_\odot$\\
    \textit{Vertical structure}\\
    $(\psi,h_c)$ & $(0.1,0.1),(0.3,0.1),(0.3,0.3)$\\
    \hline
  \end{tabular*}
  \captionsetup{width=.90\textwidth}
  \caption*{\footnotesize{$^1$: M$_{\mathrm{gas}}$ is determined from the combination of M$_{\mathrm{dust}}$ and $\Delta_{\rm gd}$.}} 
\end{table}

\subsection{Line-to-Continuum ratios}
\label{sec: line-to-continuum ratios}

As mentioned in the introduction, observing lines in the far-infrared is inherently difficult because the emission of the dust in protoplanetary disks peaks at $\sim100\ \mu\mathrm{m}$. Superimposed on this bright continuum are the intrinsically weak far-infrared lines. As a result, the line-to-continuum ratio ($L/C$) is a crucial ingredient for observing the HD levels.  

The line-to-continuum ratio can be defined as:
\begin{equation}
\label{eq: line-to-continuum}
L/C \equiv \frac{\int \mathrm{d}\nu F_{\rm line}}{\int \mathrm{d}\nu F_{\rm cont}} \simeq \frac{F_{\rm peak}}{F_{\rm cont,bin}}.
\end{equation}
Here $F_{\rm line}$ and $F_{\rm cont}$ are the continuum subtracted line flux and continuum flux, respectively. The approximation holds if the line is narrow, such that the integrated line flux can be approximated by the total flux in the center-most frequency bin ($F_{\rm peak}$) times the bin width. In this case $F_{\rm cont,bin}$ is the continuum flux in the same bin.

Due to the difficulties in characterizing the noise properties of infrared detectors \newtext{as well as cosmic ray impacts on the detector and electronics}, the current best achievable signal-to-noise ratio (SNR) for these detectors is $\sim$300. \newtext{In the most optimal case, these detectors are able to distinguish variations at the level of $\frac{1}{300} \equiv 0.0033$ of the source continuum flux.} By requiring that the HD lines are detected at the $3\sigma$ level, the SNR = 300 translates into a detection limit of $L/C$ $\geq 3\cdot0.0033 = 0.01$, which can not be improved by longer integration times.   

The spectral resolving power $R =\lambda/\Delta\lambda$ of the instrument plays a crucial role in determining the $L/C$. \newtext{As long as the line is unresolved, a higher $R$ corresponds to a larger $L/C$, independent of the disk properties. This assumption is valid for the HD lines in protoplanetary disks, which are intrinsically narrow (FWHM $\sim 5$ km s$^{-1}$ in the outer disk).} As explained in the introduction there is currently no facility capable of observing HD rotational transitions in disks. However, this work and previous studies (e.g., \citealt{Bergin2013,McClure2016}) have shown that HD is a unique tracer of the total gas mass. This means that the possibility of observing HD should be included in the consideration for instruments on future FIR observatories, such as the proposed SPICA mission. To be able to detect HD, the detection limit of $L/C\geq 0.01$ places a constraint on the requirements of the spectral resolving power of these future FIR missions.  

\begin{figure}[htb]
\centering
\includegraphics[width=0.9\columnwidth]{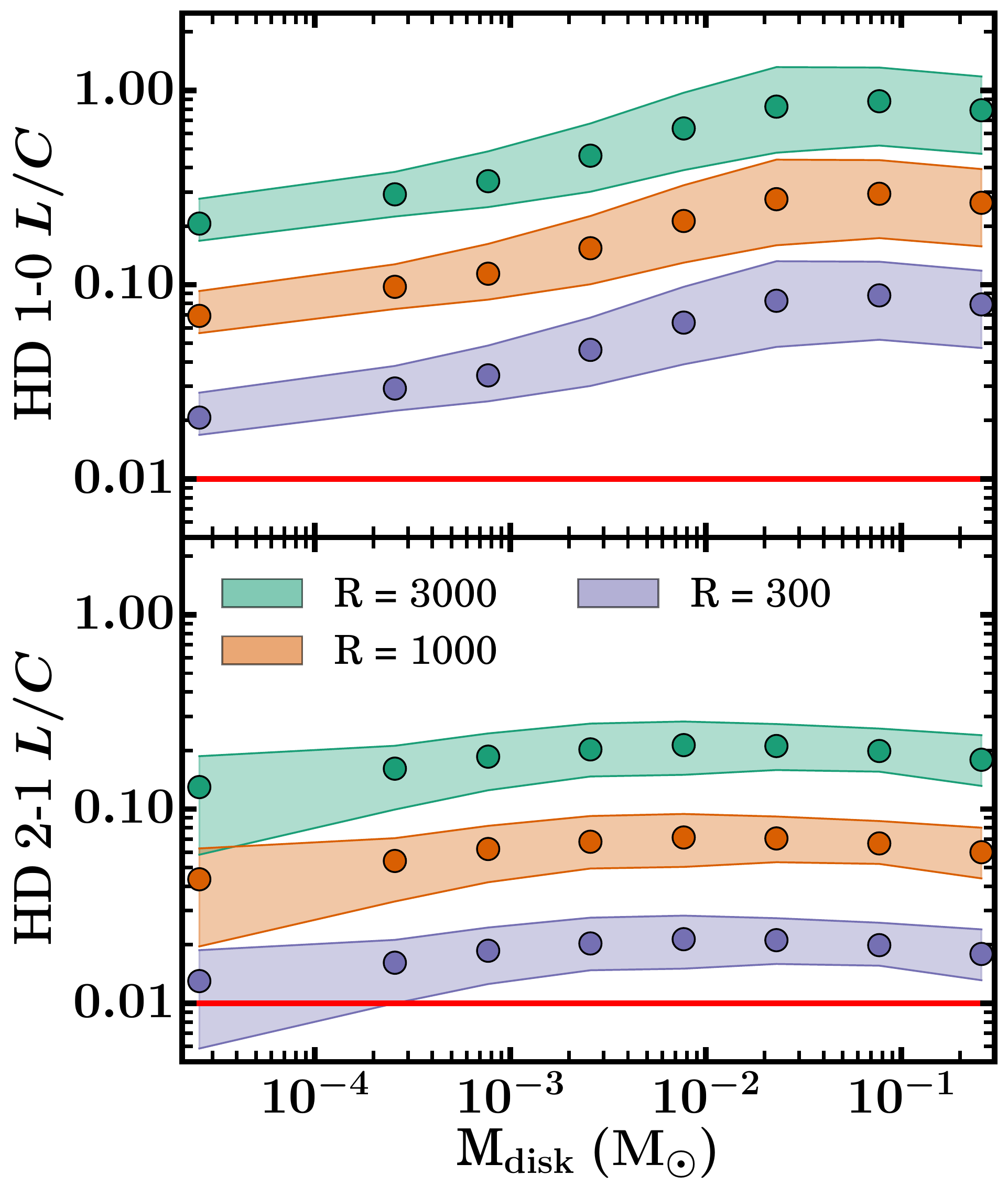}
\caption{\label{fig: LtCMass}Line-to-continuum ratios of HD as function of M$_{\rm disk}$ for different spectral resolving powers. The points show the median $L/C$ of the models in the mass bin. The shaded region shows the $L/C$s between the 16$^{\rm th}$ and 84$^{\rm th}$ percentile in each mass bin. The red line represents a $3\sigma$ detection limit of $L/C$ $=0.01$, which is equivalent to assuming a SNR on the continuum of 300.}
\end{figure}

The top panel of Figure \ref{fig: LtCMass} shows $L/C$ for HD 1-0 as function of disk mass for three different values of $R$. It is clear that the lower spectral resolving power mainly lowers the $L/C$. The current HD 1-0 detections with the \textit{Herschel} PACS instrument are for $R=1000$ at 112 $\mu$m. This figure demonstrates that future instruments, such as the SPICA SAFARI instrument \citep{Roelfsema2014}, need a resolving power of $R\geq300$ to detect HD 1-0 over the full mass range examined here, if a detection limit of $L/C\geq0.01$ (so S/N on the continuum of 300) is assumed. In the case of the HD 2-1 transition, shown in the bottom panel of Figure \ref{fig: LtCMass} a spectral resolving power of $R\geq 1000$ is required to detect HD 2-1 towards all disk masses above $10^{-5}\ \mathrm{M}_{\odot}$.

\subsection{Sensitivities of future FIR missions}
\label{sec: sensitivities}
\begin{figure}[htb]
\centering
\includegraphics[width=\columnwidth]{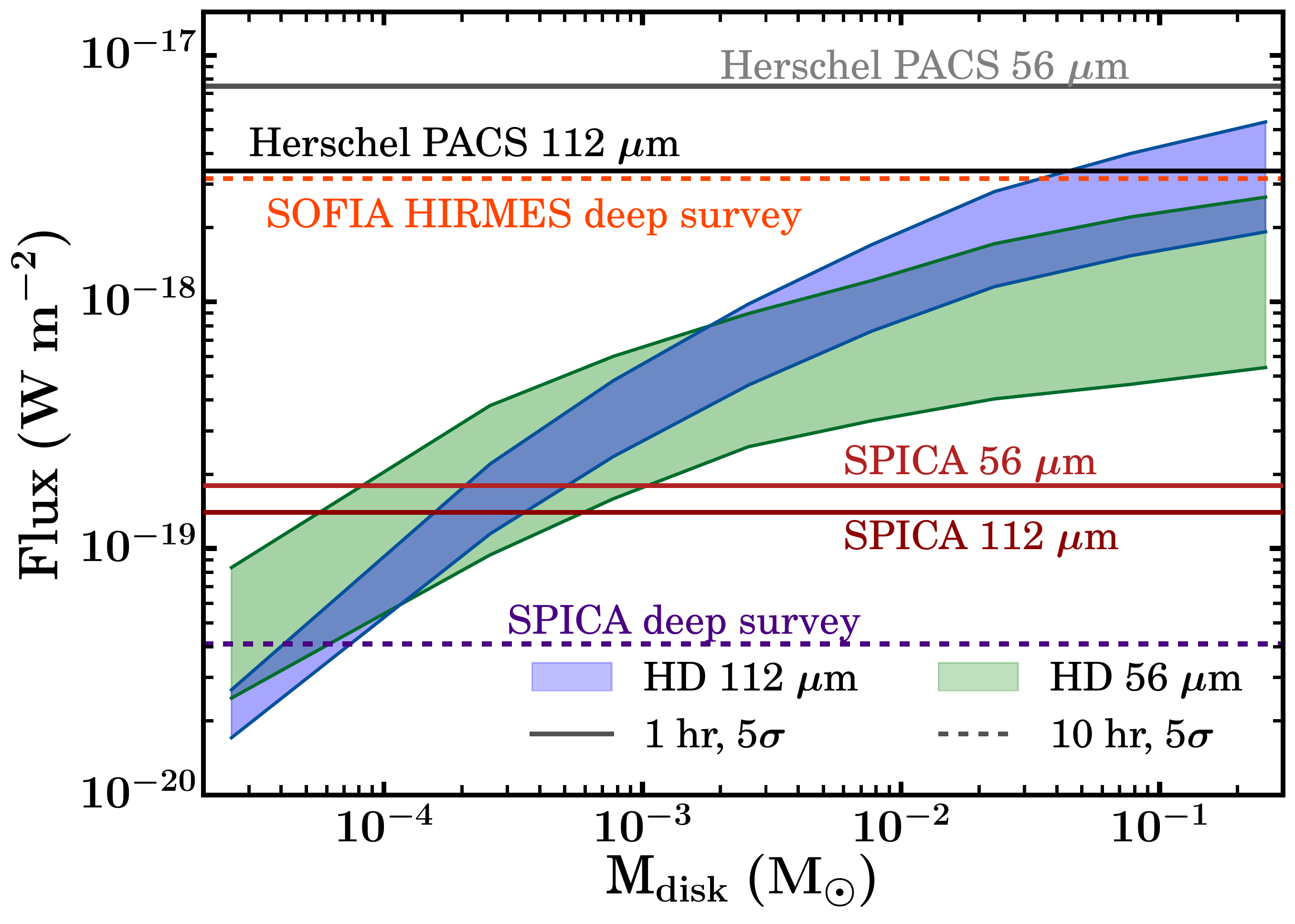}
\caption{\label{fig: sensitivity} HD 1-0 (blue) and 2-1 (green) model fluxes (calculated at 140 pc) as function of disk gas mass (cf. Figure \ref{fig: fluxMass}).  The horizontal lines represent the $5\sigma$ sensitivities for various instruments. The solid lines show the detection limit after 1 hr of integration. The dashed lines show the $5\sigma$ sensitivity at 112 $\mu$m after 10 hrs of integration (labeled `deep survey').}
\end{figure}

In addition to good line-to-continuum ratios, future FIR missions also require good sensitivities to be able to detect the HD rotational lines in protoplanetary disks. In Figure \ref{fig: sensitivity} the models presented in Figure \ref{fig: fluxMass} are compared to the sensitivities of two FIR instruments: the proposed SPICA SAFARI instrument \citep{Roelfsema2014} and the recently approved HIRMES instrument for SOFIA (E. Bergin, priv. comm.). The fluxes in the figure have been calculated for a distance of 140 pc, in line with typical distances to the closest star forming regions. The sensitivity of the Herschel PACS instrument is also shown here. The $5\sigma$ sensitivities after 1 hr of integration for SAFARI and Herschel PACS are shown in dark red and black respectively. The purple and orange lines represent the $5\sigma$ sensitivities at 112 $\mu$m after 10 hrs of integration for the SAFARI and HIRMES instruments, respectively. Figure \ref{fig: sensitivity} shows that the proposed SPICA SAFARI instrument will be able to detect HD 1-0 in disks with gas masses down to $5\cdot10^{-4}$ M$_{\odot} \approx 0.5\ \mathrm{M}_{\rm jup}$. If the integration time is increased to 10 hrs the mass sensitivity goes down to  $8\cdot10^{-5}$ M$_{\odot} \approx 0.08\ \mathrm{M}_{\rm jup}$.          


\section{Discussion}
\label{sec: discussion}

\subsection{Determining the disk gas mass}
\label{sec: determining the disk gas mass}

\newtext{In Section \ref{sec: HD flux vs. disk gas mass} the behaviour of the HD emission under variation of the disk gas mass was discussed. Figure \ref{fig: fluxMass} shows that it is possible to constrain the gas mass using the HD 1-0 flux. However, the variations in the flux due to the uncertainties on vertical structure translate into uncertainties in the determined gas mass. In this section these uncertainties are quantified and the impact of complementary observations on these uncertainties is investigated.  } 

\begin{figure}[tbh]
\centering
\includegraphics[width = \columnwidth]{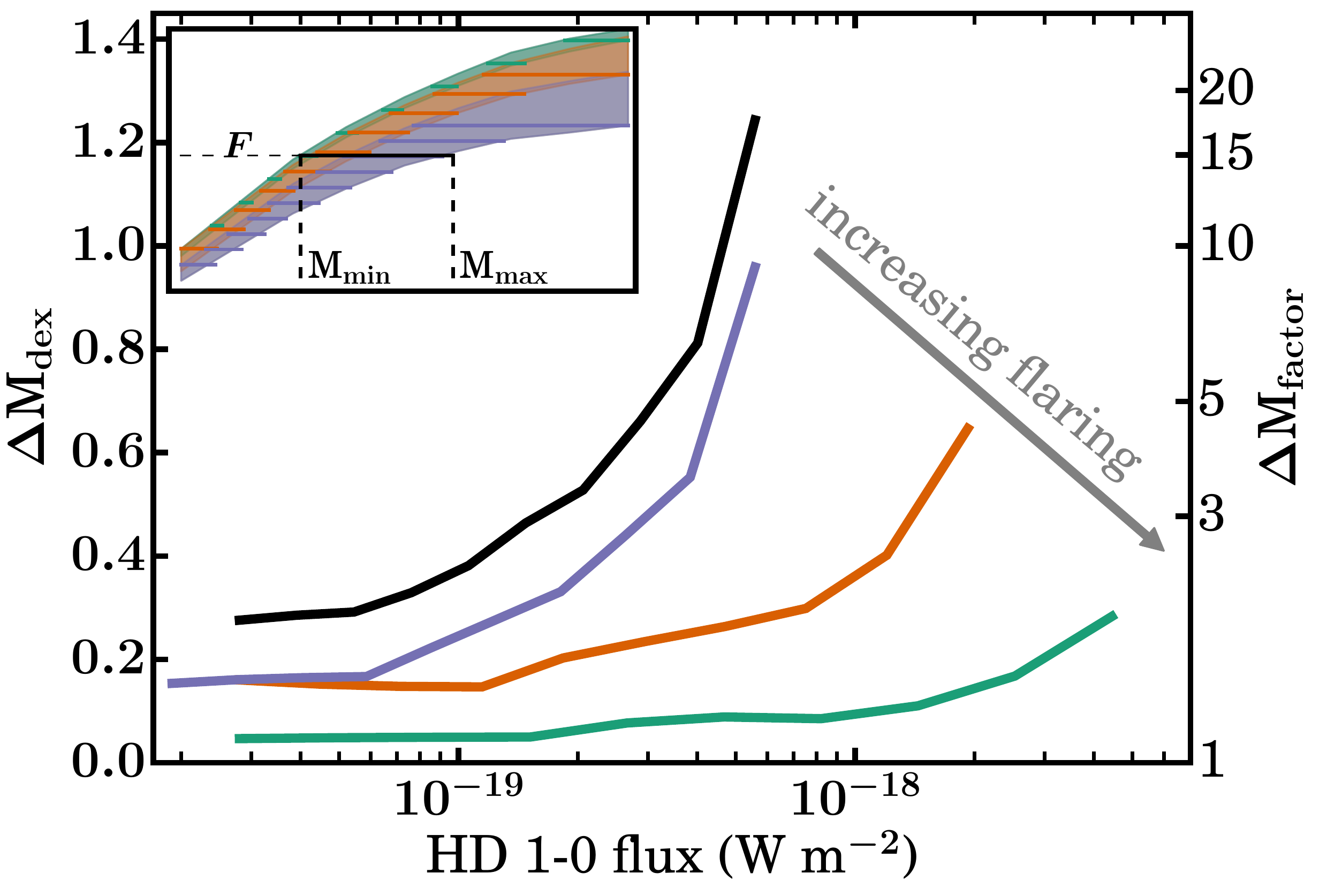}
\caption{\label{fig: massUncertainties} 
\newtext{Calculated mass uncertainties (cf. Equation \ref{eq: delta M}) as function of HD 1-0 flux. The left and right vertical axis are related via $\Delta\mathrm{M}_{\rm factor} = 10^{\Delta\mathrm{M}_{\rm dex}}$. The black line shows the result if the full set of models is used (cf. Table \ref{tab: grid parameters}). The other three lines show results for ``weakly flared'' (purple) , ``intermediately flared'' (orange) and ``strongly flared'' (light green) disks (see Table \ref{tab: vertical structure categories} for definitions for these subsets). The subsets are also shown in the top left panel, which shows a scaled version of Figure \ref{fig: fluxMass}. }} 
\end{figure}

\newtext{Figure \ref{fig: fluxMass} shows that models with different combinations of disk gas masses and vertical structures can produce the same HD 1-0 flux. The difference between the minimum and the maximum disk mass, interpolated using the model fluxes shown in Figure \ref{fig: fluxMass}, can be used to quantify the uncertainty on the gas mass estimates using the flux (cf. the small panel in Figure \ref{fig: massUncertainties}). Here the mass derivation uncertainty is defined as a factor of $10^{\Delta\mathrm{M}}$, where: }

\begin{equation}
\label{eq: delta M}
\log_{10}\Delta\mathrm{M}(F) = \frac{1}{2}\left| \log_{10}\mathrm{M}_{\rm max}(F) - \log_{10}\mathrm{M}_{\rm min}(F)\right|.
\end{equation}
\newtext{Here $\mathrm{M}_{\rm min}(F)$, $\mathrm{M}_{\rm max}(F)$ are the minimum and maximum disk masses that can produce a HD 1-0 flux $F$, given the range of vertical structures examined.}

\newtext{The calculated mass uncertainties as function of HD 1-0 flux are presented in Figure \ref{fig: massUncertainties}. For the full set of models introduced in Section \ref{sec: our models} (cf. Table \ref{tab: grid parameters}), shown in black, $\Delta$M increases with HD flux. The increase is steep, with $\Delta$M $= 1$ dex at a HD flux of $\sim 5\cdot10^{-19}$ W m$^{-2}$, indicating that this flux corresponds to an estimated mass between $10^{-3}\ \mathrm{M}_{\odot}$ and $10^{-1}\ \mathrm{M}_{\odot}$.}

This estimate relies only on the HD 1-0 flux, without making any assumptions on the vertical structure of the disk. In practice, most disks will also have been observed at other wavelengths and with other chemical tracers. These can be used to put constraints on the vertical structure of the disk. This would in turn lead to lower $\Delta$M. If complementary observations can distinguish between ``weakly flared'' disks, ``intermediately flared'' disks and ``strongly flared'' disks (see Table \ref{tab: vertical structure categories} for the definitions), the uncertainties on the mass estimates can be significantly improved.

\begin{table}[htb]
  \centering   
  \caption{\label{tab: vertical structure categories} Definition vertical structure classification}
  \begin{tabular*}{0.85\columnwidth}{lcc}
    \hline\hline
    & $\psi$ & $h_c$\\
    \hline
    ``weakly flared'' & 0.1-0.2  & 0.05-0.1  \\ 
    ``intermediately flared''   & 0.3-0.5 & 0.05-0.1\\
                                & \multicolumn{2}{c}{\textbf{or}$^1$} \\
                                & 0.1-0.2 & 0.2-0.3\\
    ``strongly flared'' & 0.3-0.5  & 0.2-0.3\\
    \hline
  \end{tabular*}
  \captionsetup{width=.8\columnwidth}
  \caption*{\footnotesize{$^1$} Note that ``intermediately flared'' disks is a joint set of two subsets of vertical structures. }
\end{table}

\newtext{The top left panel of Figure \ref{fig: massUncertainties} shows the three subsets with respect to the full set of models together with the mass uncertainties for each subset. For ``intermediately flared'' disks, shown in orange, $\Delta$M is significantly lower compared to the full set. Here the maximum mass uncertainty is $\sim0.65$ dex, meaning that the mass can be estimated within a factor of $\sim5$.}

\newtext{For the ``strongly flared'' disks, shown in light green, the uncertainty is minimized, with $\Delta$M never becoming larger than 0.2 dex. As a result, masses for this subset can be estimated to within a factor of two.}

\newtext{The mass uncertainty for the ``weakly flared'' disks follows the uncertainty of the full sample, with $\Delta$M being $\sim$0.1 dex lower than for the full sample. For these ranges of $h_c$ and $\psi$ the HD flux is quite sensitive to the vertical structure, which results in variations in the flux comparable to those in the full set. However, a difference of 0.1 dex is still a factor of two improvement on the mass uncertainty.}

\subsection{HD 1-0 and HD 2-1 line fluxes}
\label{sec: HD 1-0 and HD 2-1}
An interesting complementary observable is the HD 2-1 line emission. The main disk parameter determining the difference between the $J=1$ and $J=2$ level populations of HD is the gas temperature, which is largely set by the vertical structure. By combining observations of HD 2-1 and HD 1-0 this difference can be exploited to determine the disk mass with low uncertainty.

\begin{figure*}[htb]
\centering
\includegraphics[width=0.75\textwidth]{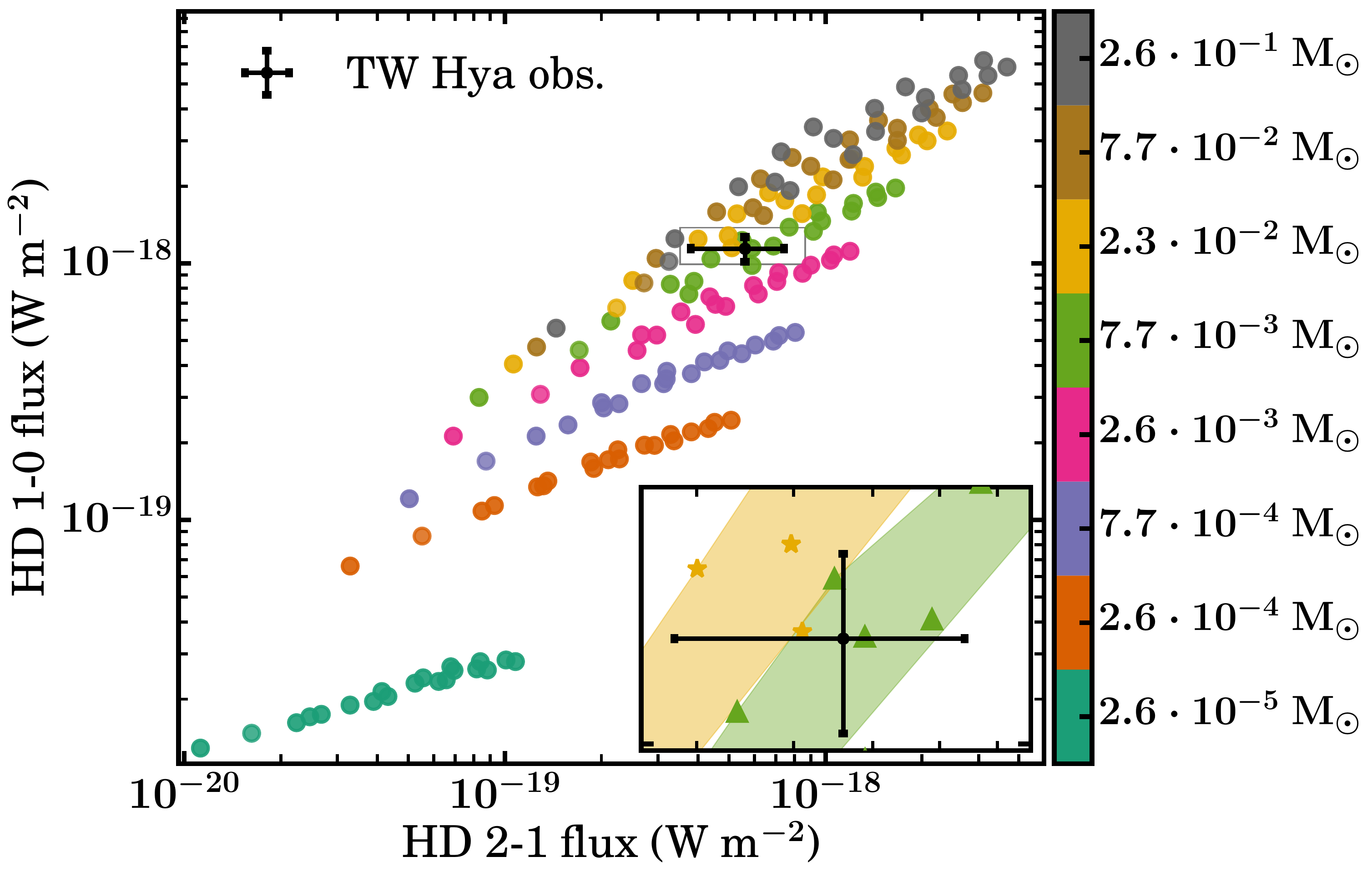}
\caption{\label{fig: comparing HD fluxes} HD 1-0 and HD 2-1 integrated fluxes (calculated at 140 pc) for the models described in Section \ref{sec: our models}. The different colours indicate different disk masses. The observations for TW Hya are shown in black (\citealt{Bergin2013,Kama2016}, D. Fedele, priv. comm.). The bottom right corner shows the region around the observations, where the shaded regions link models that are within the same mass bin.} 
\end{figure*}

As shown in Figure \ref{fig: comparing HD fluxes}, the different dependencies of HD 1-0 and HD 2-1 line fluxes on disk mass and vertical structure result in a clear mass segregation, independent of vertical structure. For disks with M$_{\rm disk} \geq 7.7\cdot10^{-3}\ \mathrm{M}_{\odot}$ the mass bins start to overlap. As the bins differ by a factor of 3 in mass, the combination of two HD lines allow the gas mass to be determined to within a factor of $\sim$3, provided that the observational uncertainties are small enough. To be able to determine the mass accurately across the whole mass range, the observational uncertainties have to decrease with increasing disk mass. This is not a too strict requirement as the observed line flux increases with disk mass, allowing for a more precise flux determination. The observations of HD 1-0 and HD 2-1 for TW Hya shown in Figure \ref{fig: comparing HD fluxes} will be discussed in Section \ref{sec: case study}.

\subsection{Comparing models to observations}
\label{sec: comparing models to observations}

\begin{figure}[htb]
\centering
\includegraphics[width=\columnwidth]{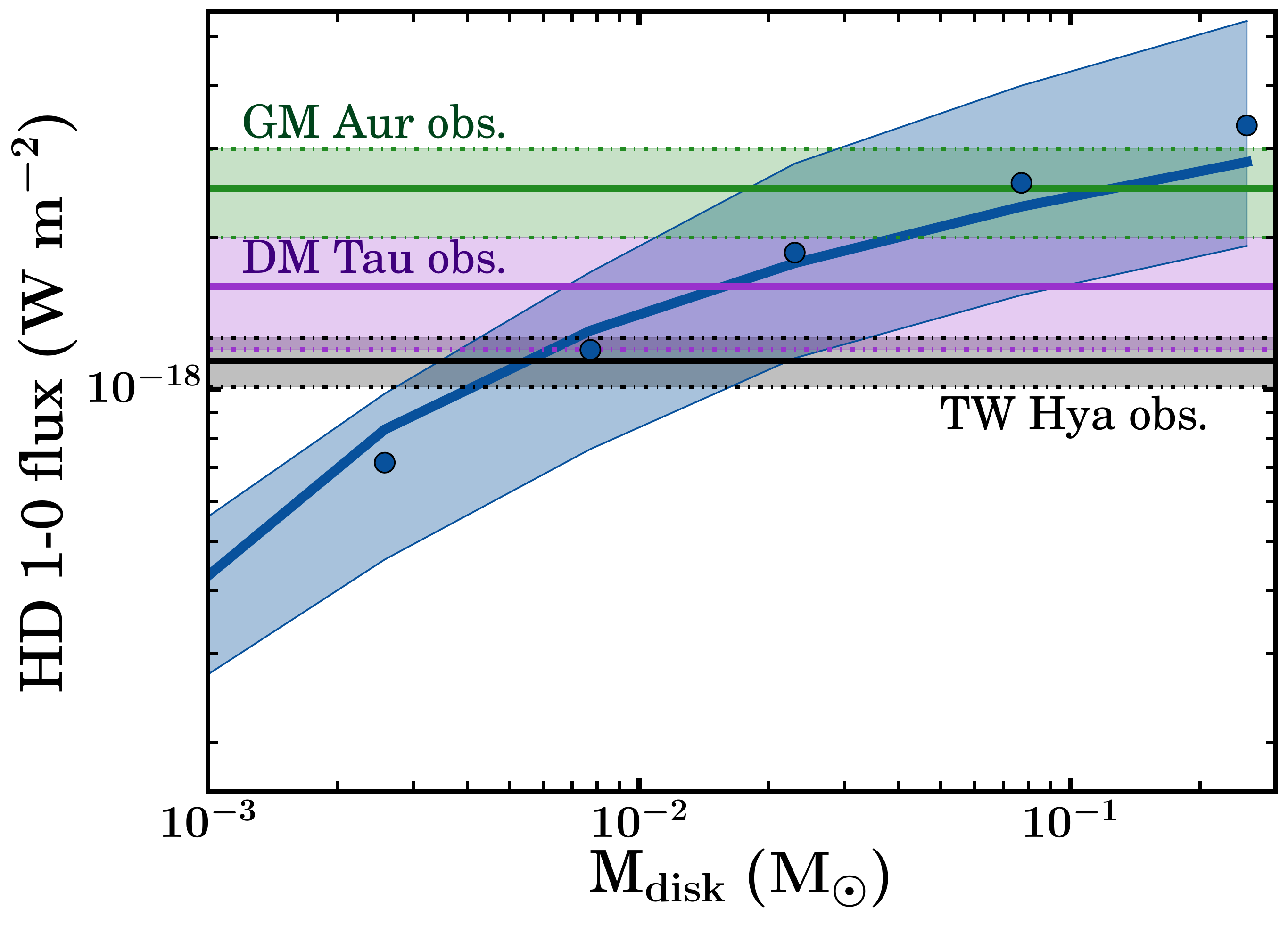}
\caption{\label{fig: observations}Integrated line flux (calculated at 140 pc) of the HD 1-0 transition as a function of disk mass, representing the high mass end of the left panel of Figure \ref{fig: fluxMass}. The horizontal black solid line gives the observation  of TW Hya, scaled to a distance of 140 pc, with the black shaded region representing the $1\sigma$ uncertainties \citep{Bergin2013}.  
The observations for DM Tau and GM Aur \citep{McClure2016} are shown in purple and green, respectively.} 
\end{figure}

The HD 1-0 transition has been observed for three disks: TW Hya \citep{Bergin2013}, DM Tau and GM Aur \citep{McClure2016}. These observations are shown in Figure \ref{fig: observations}, which represents the high mass part of the left panel of Figure \ref{fig: fluxMass}. The observed flux for TW Hya was scaled to a distance of 140 pc, to allow for comparison with the same models. 

For DM Tau the lower limit on disk gas mass based on the HD 1-0 observation is M$_{\rm disk} \approx 4\cdot10^{-3} \ \mathrm{M}_{\odot}$. This correspond to a high, strongly flared vertical structure. If instead the median vertical structure is assumed, the estimated gas mass becomes M$_{\rm disk} \approx 1.5\cdot10^{-2} \ \mathrm{M}_{\odot}$. For GM Aur, a similar analysis gives a lower limit on the disk gas mass of M$_{\rm disk} \approx 1\cdot10^{-2} \ \mathrm{M}_{\odot}$ for a high,strongly flared disk. Comparing the observed HD 1-0 flux to the medians, a gas mass of M$_{\rm disk} \approx 8\cdot10^{-2} \ \mathrm{M}_{\odot}$ is found.

Both estimates based on the medians are in line with the results of \cite{McClure2016}, who found gas masses of $(1.0-4.5)\cdot10^{-2}$ and $(2.5-19.5)\cdot10^{-2} \ \mathrm{M}_{\odot}$ for DM Tau and GM Aur by modeling both the HD flux and the continuum SED.

The observations for TW Hya shown in Figure \ref{fig: observations} will be discussed in Section \ref{sec: case study}.


\subsection{Case study: TW Hya}
\label{sec: case study}

\begin{figure}
\centering
\includegraphics[width=\columnwidth]{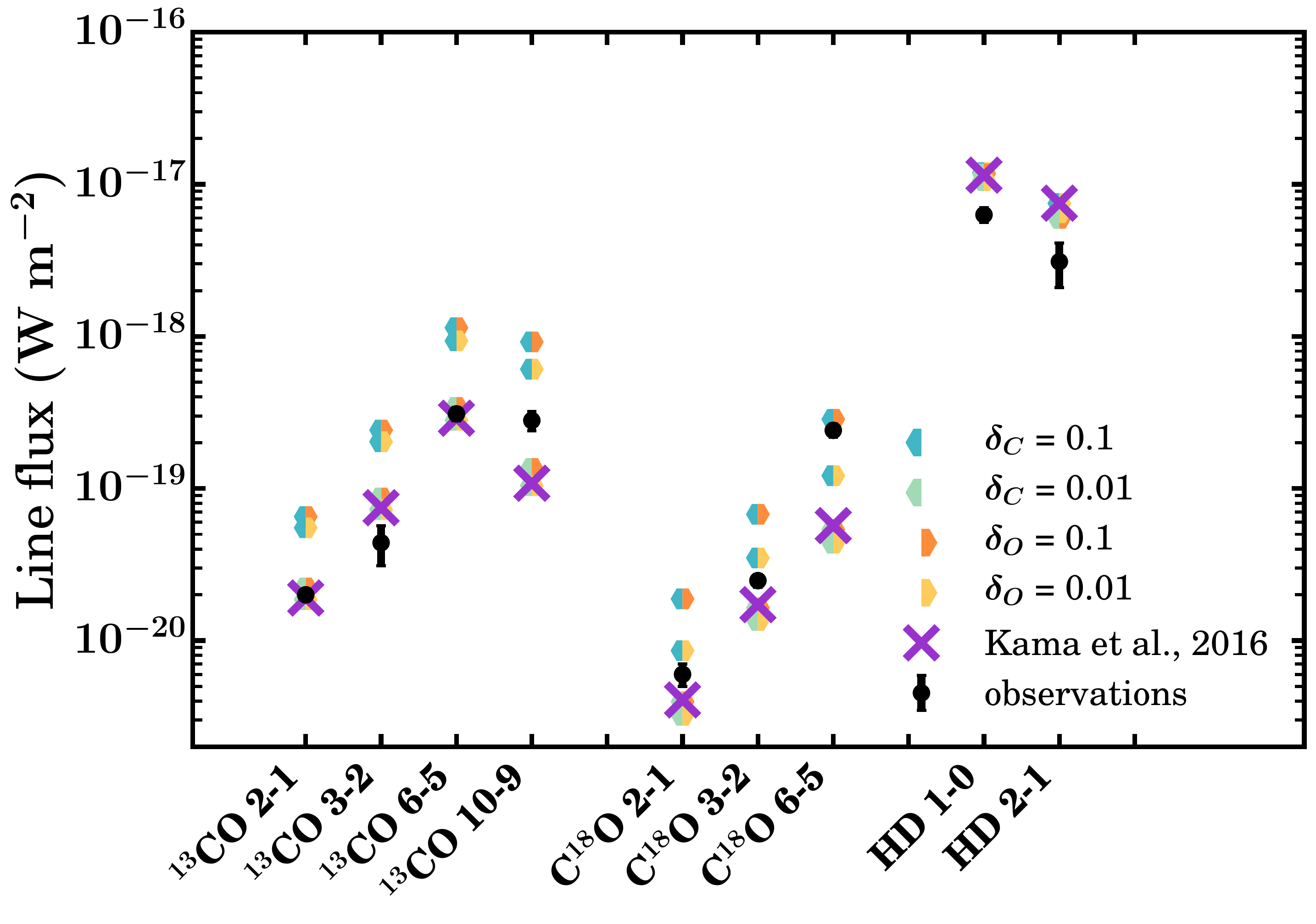}
\caption{\label{fig: LineFluxes}Integrated line fluxes from TW Hya models (M$_{\rm disk} = 2.3\cdot10^{-2}\ \mathrm{M}_{\odot}$, $\psi = 0.3$ and $h_c = 0.1$) with different amounts of carbon depletion (hexagons; left side) and oxygen depletion (right side). The black bars show the observations for TW Hya (\cite{Schwarz2016,Kama2016} and references therein). The purple crosses show the results from \cite{Kama2016}, recalculated for $\delta_{\rm C} = \delta_{\rm O} = 0.01$, [D]/[H] = $2\cdot10^{-5}$ (cf. Section \ref{sec: case study}).}
\end{figure}

The protoplanetary disk of TW Hya ($59.54\pm 1.45$ pc, \citealt{AstraatmadjaBailerJones2016}), one of the best-studied disks. Its disk gas mass has been estimated to be $\geq 0.05\ \mathrm{M}_{\odot}$ using HD \citep{Bergin2013}. This estimate is an order of magnitude higher than a similar estimate of the gas mass using C$^{18}$O \citep{Favre2013}.

The observations of the HD 1-0 and 2-1 line fluxes for the TW Hya disk are presented in Figures \ref{fig: comparing HD fluxes} and \ref{fig: observations}. Comparing these observations to the models shows that the observed HD 1-0 flux for TW Hya places a lower limit of M$_{\rm disk} \approx 3\cdot 10^{-3} \ \mathrm{M}_{\odot}$ if high, strongly flared vertical structure is assumed. Using the radial and vertical structure from \cite{Kama2016}, the observations constrain the disk gas mass to $6\cdot10^{-3} \ \mathrm{M}_{\odot} \leq \mathrm{M}_{\rm disk} \leq 9\cdot10^{-3} \ \mathrm{M}_{\odot}$. A similar gas mass is found from combining the observations of HD 1-0 and HD 2-1 (cf. Figure \ref{fig: comparing HD fluxes}), which favor a gas mass between $7.7\cdot10^{-3}\ \mathrm{M}_{\odot}\leq\mathrm{M}_{\rm disk}\leq2.3\cdot10^{-2}$ M$_{\odot}$. Both estimates are lower than previous estimates (e.g., \citealt{Bergin2013,Kama2016}). Note that the estimates lie closer to the disk mass estimated from C$^{18}$O observations \citep{Favre2013}, suggesting a smaller tension between the two. 

Using models that include detailed modeling of the CO and H$_2$O chemistry, studies found that the disk surrounding TW Hya is genuinely depleted in volatile carbon and oxygen (e.g.,  \citealt{Bergin2010,Bergin2013,Hogerheijde2011,Favre2013,Du2015}). Recent observations of [C{\small I}], [C{\small II}], [O{\small I}], C$_2$H and C$_3$H$_2$ have reinforced this conclusion \citep{Kama2016,Bergin2016}. 

In their model, \cite{Kama2016} incorporate the isotopes $^{13}$C, $^{17}$O, $^{18}$O and D  parametrically. In this section, their best fitting model (M$_{\rm disk} = 2.3\cdot10^{-2}\ \mathrm{M}_{\odot}$, $\psi = 0.3$ and $h_c = 0.1$) is rerun using the extended chemical network that includes the deuterium and CO isotopologue selective chemistry following \cite{Miotello2014}. For these models, the carbon abundance was varied between the ISM abundance and two consecutive orders of underabundance, i.e., X(C) $\equiv$ [C]/[H] $\in [1.35\cdot10^{-4},\ 1.35\cdot10^{-5},\ 1.35\cdot10^{-6}]$. The same was done for the oxygen abundance, X(O) $\in [2.88\cdot10^{-4},\ 2.88\cdot10^{-5},\ 2.88\cdot10^{-6}]$. These underabundances are denoted using $\delta_i \equiv$ X($i$)/X($i$)$_{\rm ISM}$. \newtext{Note that in order to match the hydrocarbon abundances \cite{Kama2016} fine-tuned their final model to $\delta_{\rm C} = 0.01$ and C/O = 1.5, while the disk structure was unchanged from the best-fit fiducial model.}

Figure \ref{fig: LineFluxes} shows the resulting line fluxes for several rotational transitions of HD, $^{13}$CO and C$^{18}$O, similar to Figure 6 in \cite{Kama2016}. In the figure only the models with $\mathrm{X(C,O)} \sim 10^{-6}, 10^{-5}$ are shown. The full figure can be found in Appendix \ref{app: line fluxes TW Hya model}. It should be noted that HD abundance used in \cite{Kama2016} ($n(\mathrm{HD})/n(\mathrm{H}_2) = 1\cdot10^{-5}$) is a factor four lower than the maximal HD abundance of the models in this work ($n(\mathrm{HD})/n(\mathrm{H}_2) = 4\cdot10^{-5}$). In order to compare the output of the chemical calculations to their parametric method, the HD line fluxes of their model were recalculated using the higher HD abundance. This accounts for the differences in the HD line fluxes seen here and those in \cite{Kama2016}. Similar steps where taken so that the $\delta_{\rm C} = \delta_{O} = 0.01$ model and the \cite{Kama2016} model have the same oxygen and carbon abundances. 

Starting with HD, one can see that the line fluxes from the models run here, which include the deuterium chemistry network, match the line fluxes from the model with the parametric HD within a few percent (8\% and 15\% for HD 1-0 and HD 2-1, respectively). 
This can be understood from the results in Section \ref{sec: HD emission maps}, which show that all the available deuterium is locked up in HD, thus mimicking the parametric approach used by \cite{Kama2016}. 
Note that all models overproduce the observed HD fluxes by a factor $\sim2$, suggesting a slightly lower disk gas mass for TW Hya than used by \cite{Kama2016}. 

For $^{13}$C, there is again good agreement between the fluxes from \cite{Kama2016} and the $\delta_{\rm C} = \delta_{O} = 0.01$ model (its CO isotopologue counterpart). This can be understood by reactions such as the ion-molecule reaction $^{13}$C$^{+}$ + $^{12}$CO $\rightleftharpoons$ $^{13}$CO + $^{12}$C$^{+}$ + 35 K. The forward reaction direction of this reaction is favoured at low temperatures, leading to an increased $^{13}$CO abundance which balances out the additional decrease of $^{13}$CO due to isotope-selective photodissociation \citep{Miotello2014}.

The line fluxes of C$^{18}$O of the $\delta_{\rm C} = \delta_{O} = 0.01$ model are systematically lower than the same lines from \cite{Kama2016}. This qualitatively agrees with previous results (e.g., \citealt{Visser2009,Miotello2014}), but the effect is smaller than has been suggested. 
A possible explanation can be found in the gas mass of the disk. \cite{Miotello2014,Miotello2016} show that for disks with M$_{\rm disk} \geq 7\cdot10^{-3}\ \mathrm{M}_{\odot}$ freeze-out is the dominant process affecting the C$^{18}$O flux, whereas isotope-selective photodissociation is more important for low mass disks (see, e.g., Fig. 3 in \citealt{Miotello2016}).

When the models are compared to the observations to determine the carbon and oxygen underabundance the results vary based on the tracer used. The $^{13}$CO observations point to $\delta_{\rm C}$ = 0.01, in line with the value found by \cite{Kama2016}. The observed $^{13}$CO 10-9 flux and the observations for C$^{18}$O instead favour a lower underabundance of carbon, between $\delta_{\rm C}$ = 0.01 and $\delta_{\rm C}$ = 0.1. For all CO isotopologue lines $\delta_{\rm O}$ = 0.01 and $\delta_{\rm O}$ = 0.1 fit the observations equally well. Combined with the lower gas mass estimates found for TW Hya, these results show that the inclusion of isotope-selective processes decreases underabundance of carbon needed to explain the observations to $\delta_{\rm C}\approx0.1$. It should be noted however that in this analysis only the CO isotopologues are considered, whereas the analysis in \cite{Kama2016} also included observations of atomic carbon and atomic oxygen. 


\section{Conclusions}
\label{sec: conclusions}
Accurately determining the amount of material in protoplanetary disks is crucial for understanding both the evolution of disks and the formation of planets in these disks. Recent observations of the rotational transitions of HD in protoplanetary disks \citep{Bergin2013,McClure2016} have added a new possibility of tracing the disk gas mass. In this work, the robustness of HD as a tracer of the disk gas mass and its dependence on the vertical structure are investigated. The thermochemical code \texttt{DALI}  \citep{Bruderer2012,Bruderer2013} was used to calculate the line fluxes for the disk models. The normal chemical network of \texttt{DALI}  was expanded to include CO isotopologue selective chemistry, following \cite{Miotello2014}. A simple D chemistry network was also implemented in \texttt{DALI}. A series of models was run spreading a range of disk masses (M$_{\rm disk} \sim10^{-5}-10^{-1}\ \mathrm{M}_{\odot}$) and vertical structures ($h_c \sim 0.05-0.3,\ \psi \sim 0.1-0.5$), with the large grains settled close to the midplane. From these models, the following observations can be made:
\begin{itemize}
    \setlength\itemsep{1em}
    \item The HD flux increases as a powerlaw for low mass disks (M$_{\rm disk} \leq 10^{-3}$ M$_{\odot}$). The fitted powerlaw indices are 0.8 for HD 1-0 (112 $\mu$m) and 0.5 for HD 2-1 (56 $\mu$m). These indices are less than 1.0 due to the dependence of the emission on the gas temperature. For high mass disks (M$_{\rm disk} > 10^{-3}$ M$_{\odot}$) the HD flux scales with $\log_{10}\ \mathrm{M}_{\rm disk}$, which is a result of the increased line optical depth and decreased overall temperature of these disks.
    \item The maximum variation in HD 1-0 flux due to the different vertical structure increases with disk mass, from $0.75\ \times$ the median flux for $\mathrm{M}_{\rm disk} \sim 10^{-5}\ \mathrm{M}_{\odot}$ up to $1.9\ \times$ the median flux for $\mathrm{M}_{\rm disk} \sim 10^{-1}\ \mathrm{M}_{\odot}$. The variation for HD 2-1 is both larger than for HD 1-0 and is less dependent on disk mass. The maximum variation of this flux is $1.9\ \times$ the median flux for $\mathrm{M}_{\rm disk} \sim 10^{-5}\ \mathrm{M}_{\odot}$, increasing to $2.8\ \times$ the median flux for $\mathrm{M}_{\rm disk} \sim 10^{-1}\ \mathrm{M}_{\odot}$. 
    \item The influence of the large grain population on the HD flux is less than that of the vertical structure, approximately 0.7$\times$ the median flux at M$_{\rm disk} = 2.3\cdot10^{-2} \ \mathrm{M}_{\odot}$.
    \item Without making assumptions on the vertical structure of the disk, the HD 1-0 flux can be used to estimate the disk gas mass to within a factor of \newtext{$\sim$3 for low mass disks (M$_{\rm disk} \leq10^{-3}$ M$_{\odot}$)}. For more massive disks the uncertainty in the estimated mass increases to more than an order of magnitude.
    \item \newtext{If complementary observations are employed to constrain the disk's flaring, the uncertainty on the gas mass can be reduced down to factor of 2, even for massive disks. Moreover, a combination of HD 1-0 and the HD 2-1 fluxes can be used to a determine the disk gas mass to within a factor of 3, without making assumptions on the vertical structure}.
    \item For DM Tau and GM Aur, gas mass estimates found by comparing the observed fluxes to the models agree with the results from \cite{McClure2016}, confirming the high gas masses of these disks.
    \item  For TW Hya, a gas mass between $6\cdot10^{-3} \ \mathrm{M}_{\odot} \leq \mathrm{M}_{\rm disk} \leq 9\cdot10^{-3} \ \mathrm{M}_{\odot}$ is found if the best fit vertical structure from \citep{Kama2016} is assumed. This estimate agrees with the combination of HD 1-0 and HD 2-1 line fluxes, which favour a gas mass of $7.7\cdot10^{-3}\ \mathrm{M}_{\odot}$. 
    \item Detailed modeling of the TW Hya disk shows that the difference between HD- and CO-based gas masses is mitigated by including CO isotopologue selective effects. A carbon underabundance of $\sim10$ with respect to the ISM can also explain the CO isotopologue observations.
\end{itemize}
In the interest of future far-infrared observatories, the line-to-continuum ratios for HD 1-0 and HD 2-1 where calculated for different spectral resolving power. If a maximum signal-to-noise on the continuum of 300 is assumed, it was shown that future far-IR missions need a spectral resolving power $R \geq 300$ (equivalent to $\Delta v \leq 500$ km s$^{-1}$) to detect HD 1-0 for all disk masses. To detect HD 2-1 towards all models, a spectral resolving power of $R \geq 1000$ (equivalent to $\Delta v \leq 150$ km s$^{-1}$) is required. Furthermore, a $5\sigma$ sensitivity of $1.8\cdot10^{-20}$ W m$^{-2}$ ($2.5\cdot10^{-20}$ W m$^{-2}$) is required to detect HD 1-0 (2-1) in disks with gas mass $\geq 3\cdot10^{-5}\ \mathrm{M}_{\odot}\approx0.03\ \mathrm{M}_{\rm jup}$. \newtext{Both requirements can be met by proposed future missions such as SPICA.}

\begin{acknowledgements}
The authors would like to thank Inga Kamp and Ted Bergin for the useful discussions and for providing the details for the SPICA SAFARI and SOFIA HIRMES instruments. They also acknowledge Arthur Bosman for his support with the DALI code and Davide Fedele for providing the HD 2-1 line flux that was observed for TW Hya. Astrochemistry in Leiden is supported by the Netherlands Research School for Astronomy, by a Royal Netherlands Acadamy of Arts and Sciences (KNAW) professor prize, and by the European Union A-ERC grant 291141 CHEMPLAN.
\end{acknowledgements}

\bibliographystyle{aa} 
\bibliography{references}

\begin{appendix}
\section{Abundance and temperature maps of TW Hya}
\begin{figure}
    \begin{subfigure}{\columnwidth}
        \includegraphics[width=\columnwidth]{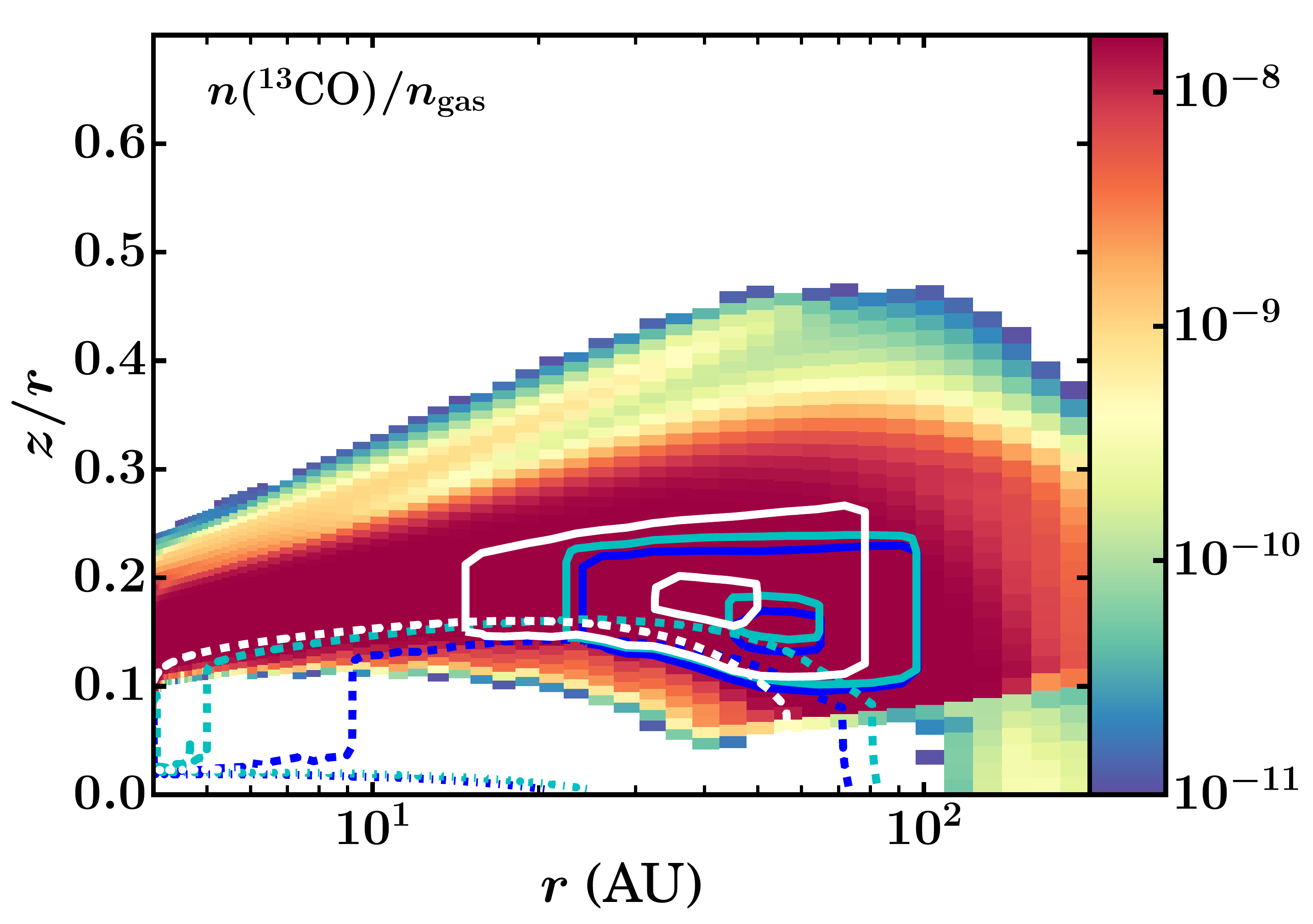}
    \end{subfigure} 
    \begin{subfigure}{\columnwidth}
        \includegraphics[width=\columnwidth]{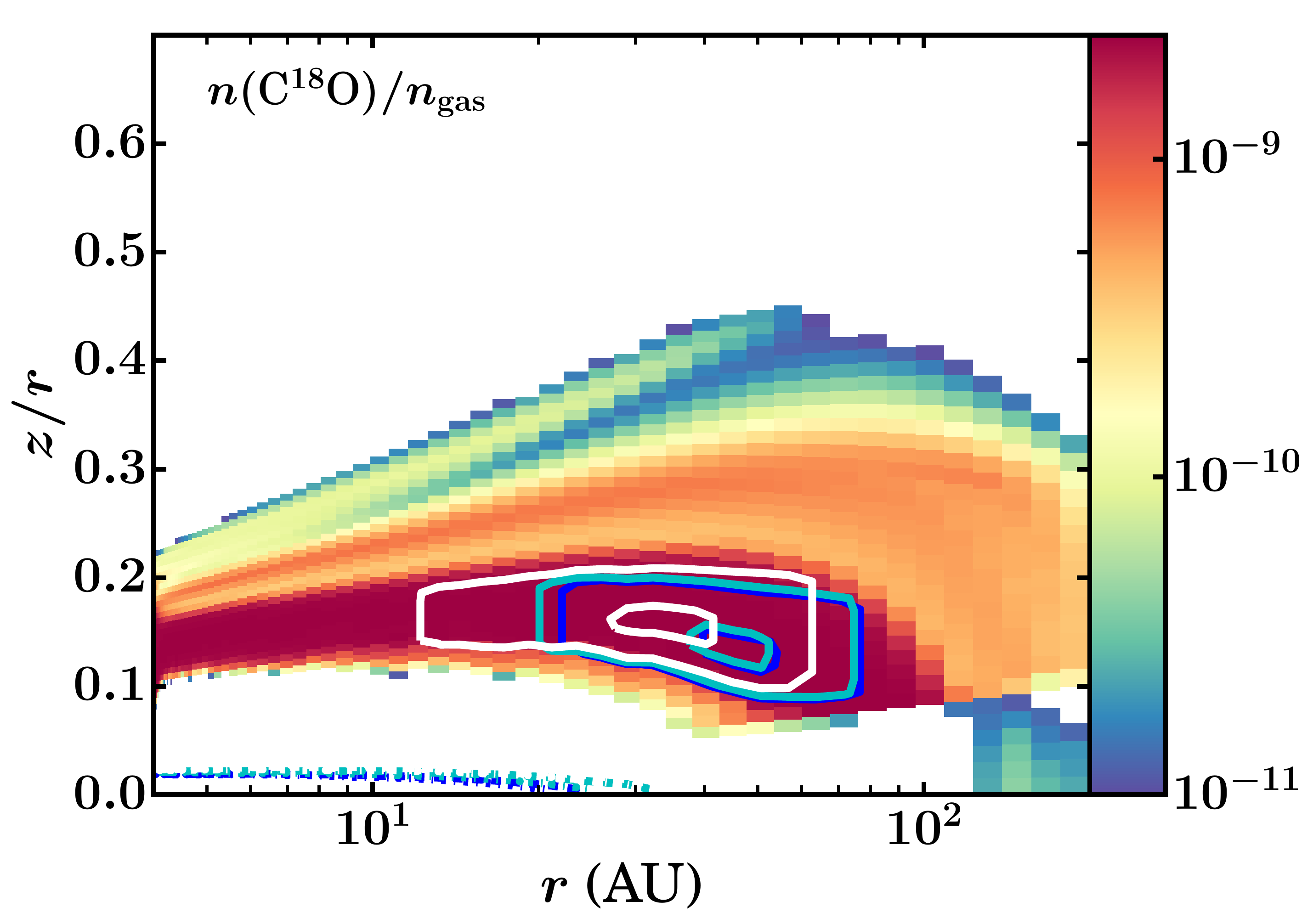}
    \end{subfigure}
    \caption{\label{fig: CO isotopologues}Maps of the $^{13}$CO (top) and C$^{18}$O (bottom) abundances. The coloured lines correspond to $^{x}$C$^{y}$O 2-1 (blue), 3-2 (light blue) and 6-5 (white) respectively. The solid lines denote where 25\% and 75\% of the emission is produced. the dashed lines. The dashed lines show the $\tau=1$ surface for the line emission. The dashed-dotted lines show the $\tau=1$ surface for the continuum at the wavelength of the line. }
\end{figure}

\begin{figure}
    \begin{subfigure}{0.95\columnwidth}
        \includegraphics[width=\columnwidth]{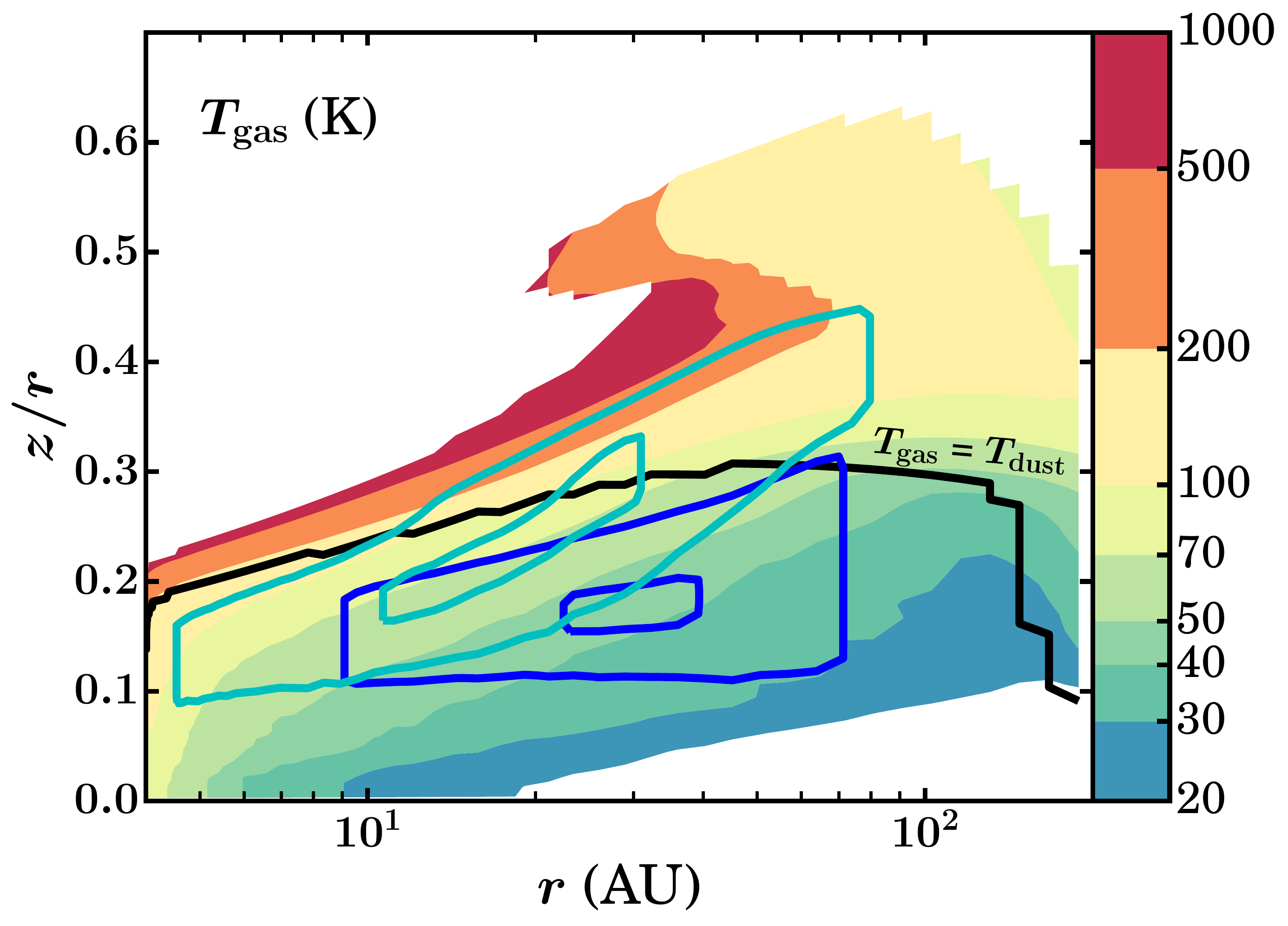}
    \end{subfigure}
    \qquad
    \begin{subfigure}{0.95\columnwidth}
        \includegraphics[width=\columnwidth]{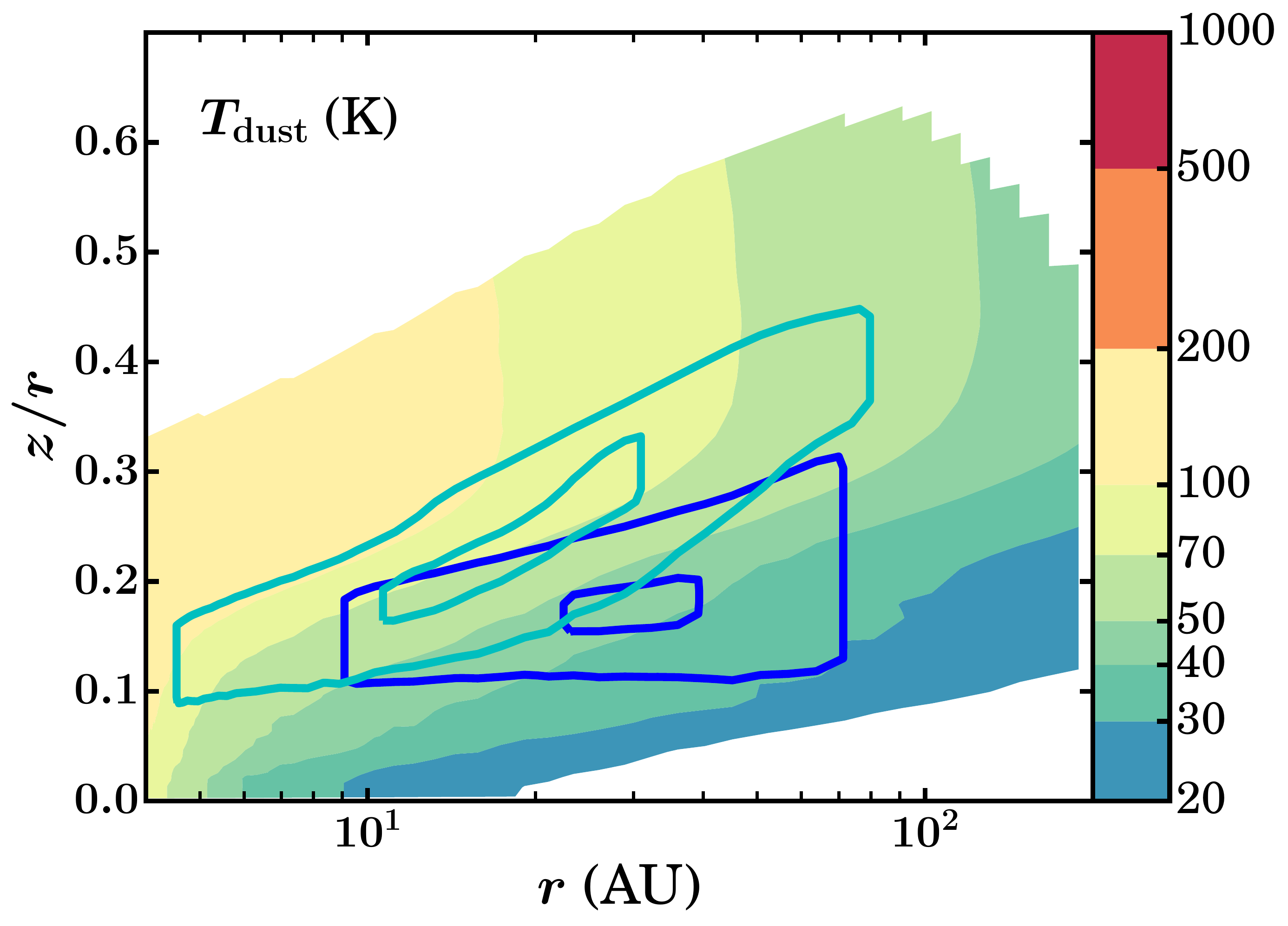}
    \end{subfigure}
    \caption{\label{fig: Temperatures} Maps of the gas temperature (top), dust temperature (bottom).
    The solid lines denote where 25\% and 75\% of the emission from HD 1-0 (blue) and HD 2-1 (light blue) originate from. The region below the black line in the top panel is where $T_{\rm gas} = T_{\rm dust}$. }
\end{figure}

In Figure \ref{fig: CO isotopologues} the abundance maps of $^{13}$CO and C$^{18}$O are shown for the TW Hya model (M$_{\rm disk} = 2.3\cdot10^{-2}\ \mathrm{M}_{\odot}$, $h_c = 0.1$, $\psi = 0.3$ and  $\delta_{\rm C} = \delta_{\rm O} = 0.01$.) The gas and dust temperature structures for the same model are shown in the panels of Figure \ref{fig: Temperatures}. 

\section{Abundance and emission maps of grid models}
\label{app: emission maps grid models}
\begin{figure}
    \includegraphics[width=\columnwidth]{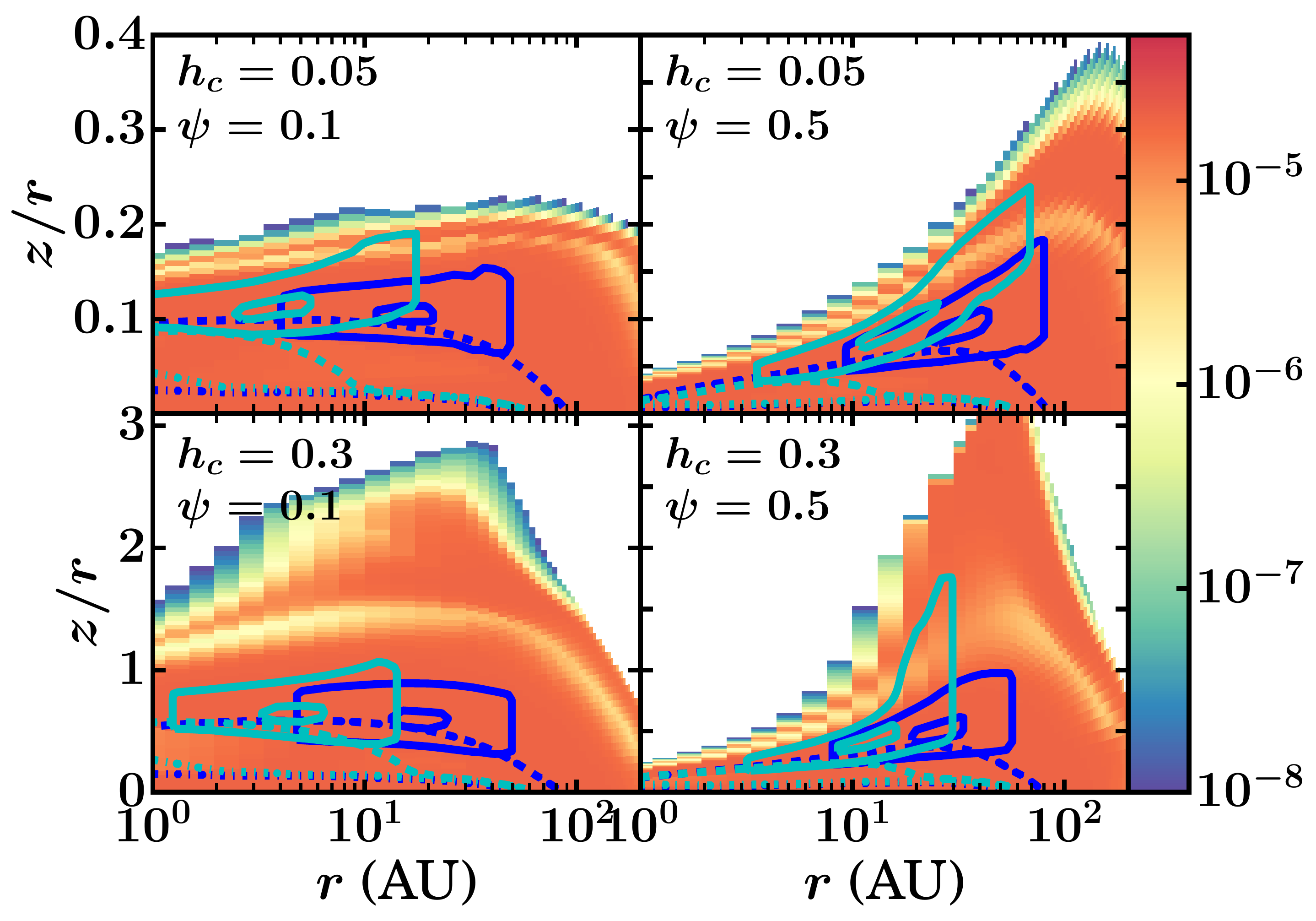}
    \caption{\label{fig: emission vertical structure} Maps of the HD abundance for four models with different vertical structures. Starting with the top left figure and continuing clockwise the vertical structure shown are $(h_c, \psi) = [0.05,0.1], [0.05,0.5], [0.3,0.1], [0.3,0.5]$. Note that the top and bottom row have different ranges on the vertical axis. The solid lines denote where 25\% and 75\% of the emission from HD 1-0 (blue) and HD 2-1 (light blue) originate from.}
\end{figure}

The HD abundance maps for several models with different vertical structures and disk masses are shown here. Figure \ref{fig: emission vertical structure} presents four models with different vertical structures: $(h_c, \psi) = [0.05,0.1], [0.05,0.5], [0.3,0.1], [0.3,0.5]$. These represent the most extreme cases in the grid. Note that the top and bottom rows in the figure have different vertical extents. Also note that the HD 1-0 emission (blue solid contours) originates from approximately the same horizontal region in all four models, between a few AU and several tens of AU. 

\begin{figure}
    \includegraphics[width=0.95\columnwidth]{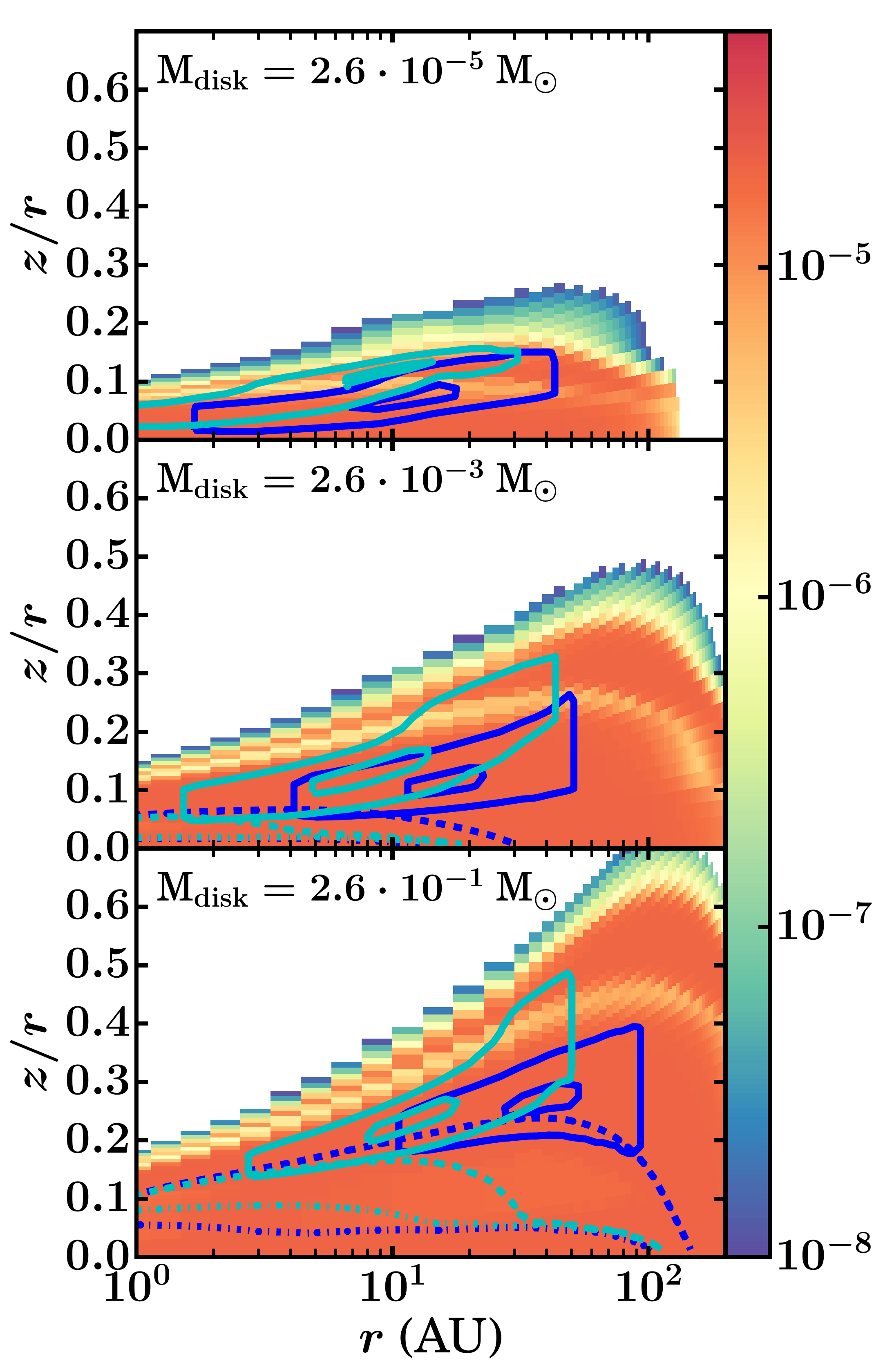}
    \caption{\label{fig: emission mass} Maps of the HD abundance for three models with different disk masses: M$_{\rm disk} = [10^{-5}, 10^{-3}, 10^{-1}]\ \mathrm{M}_{\odot}$. The solid lines denote where 25\% and 75\% of the emission from HD 1-0 (blue) and HD 2-1 (light blue) originate from. }
\end{figure}

Figure \ref{fig: emission mass} presents the abundance maps for three disks with different disk masses: M$_{\rm disk} = [10^{-5}, 10^{-3}, 10^{-1}]\ \mathrm{M}_{\odot}$. Note that the height $\tau = 1$ surface for the line optical depth (blue dashed line) increases with disk mass.

\section{Effects of including hydrostatic equilibrium}
\label{app: effects of including hydrostatic equilibrium}

\newtext{As shown in Figure \ref{fig: HDdistribution}, the HD emission originates from regions close to the disk surface. Accordingly, the HD emission will depend on the assumed disk vertical structure. In the models presented here the vertical structure is parametrized with a Gaussian (cf. Equation \ref{eq: vertical density}), to mimic a balance between gravitational force and pressure gradient, which in reality would depend on the temperature structure of the disk. Another possible approach is to the hydrostatic equilibrium to derive the vertical structure (e.g., \citealt{Woitke2009}). This results in a puffed-up inner rim, which shadows the outer disk from direct irradiation by the star. However, this may be a temporary feature that could be dispersed by e.g. stellar winds.}

\newtext{For completeness, we have investigated how the inclusion of hydrostatic equilibrium would affect the HD emission. The fiducial model described in Section \ref{sec: HD emission maps} was rerun using eight iterations of the hydrostatic solver of DALI, where the temperature structure of the $(n-1)^{\rm th}$ model is used to calculate the vertical structure of the $n^{\rm th}$ model. The resulting vertical structure is then used to recalculate the dust and gas temperatures, the chemistry and the HD emissions.}

\subsection{The hydrostatic solver}
\label{app: hydrostat solver}

\newtext{One way to obtain the vertical structure of the disk is to solve the equations of hydrostatic equilibrium. The hydrostatic equations in cylindrical geometry read (e.g. \citealt{hartmann2000})}

\begin{align}
\label{eq: hydrostatic equations}
\rho \frac{v_\phi}{r}   &= \frac{\mathrm{d}P}{\mathrm{d}r} + \rho \frac{\mathrm{d}\phi}{\mathrm{d}r}\\
                    0   &= \frac{\mathrm{d}P}{\mathrm{d}z} + \rho \frac{\mathrm{d}P}{\mathrm{d}z},   
\end{align}
\newtext{where the velocities in the $z$ and $r$ direction are assumed to be 0. Without self-gravity, the gravitational potential by the star is}
\begin{equation}
\label{eq: grav pot}
\phi = -\frac{GM_*}{\left(r^2 + z^2\right)^{1/2}},
\end{equation}
where $M_*$ is the mass of the star.

\newtext{Assuming that the pressure gradient in the $r$ direction is negligible, the first hydrostatic equation yields the Keplerian velocity profile and the second hydrostatic equation can be solved independently for each radius. Instead of directly solving the equation for the density $\rho$, the equation can also be solved for pressure. The pressure and the density are related through $P = c_s^2\rho$, where $c_s$ is the sound speed which depends on the gas temperature $T_{\rm gas}$ and therefore varies with height $z$. The differential equation for $P$ reads}
\begin{align}
\label{eq: differential P}
\frac{\mathrm{d}P}{\mathrm{d}z} &= -\rho\frac{\mathrm{d}\phi}{\mathrm{d}z}\\
\frac{1}{P} \frac{\mathrm{d}P}{\mathrm{d}z} &= - \frac{zGM_*}{c_s^2 \left(r^2 + z^2\right)^{3/2}},
\end{align}

\noindent \newtext{which is solved by}
\begin{equation}
\label{eq: P solution}
P(z) = c_s(z)^2\rho(z) = C \exp\left( \int \frac{zGM_*}{c_s^2 \left(r^2 + z^2\right)^{3/2}}\mathrm{d}z\right).
\end{equation}

\newtext{In the case of an isothermal disk, the above equation reduces to a Gaussian vertical density profile, as given in equation \eqref{eq: vertical density}. The integration constant $C$ in the above equation results from the normalization conditions for the vertical column density and should thus adjusted such that $\int \rho(z)\mathrm{d}z$ is preserved.}

\subsection{Comparing with parametrized vertical structure}

\begin{figure}
    \begin{subfigure}{0.95\columnwidth}
        \includegraphics[width=\columnwidth]{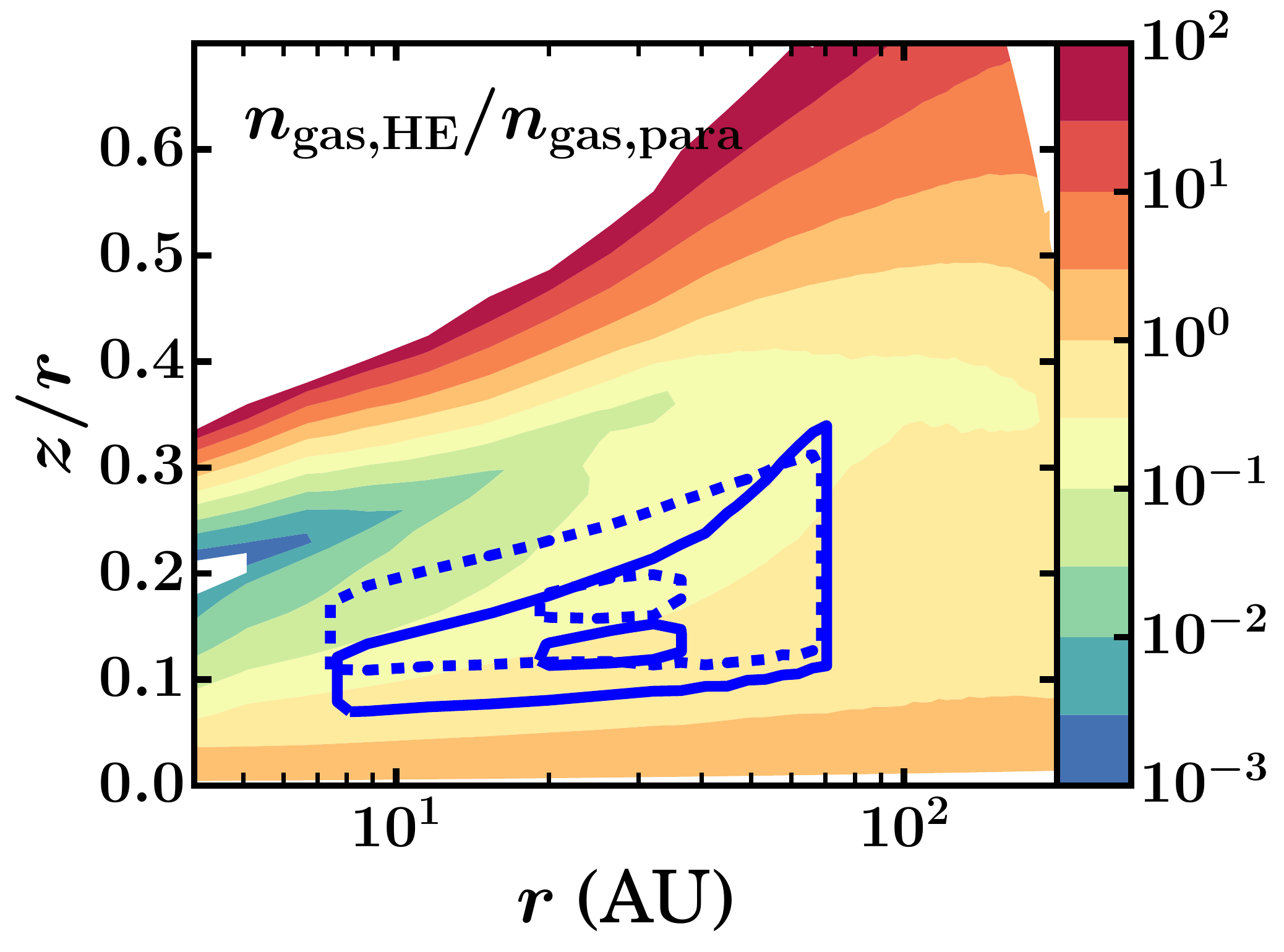}
    \end{subfigure}
    \qquad
    \begin{subfigure}{0.95\columnwidth}
        \includegraphics[width=\columnwidth]{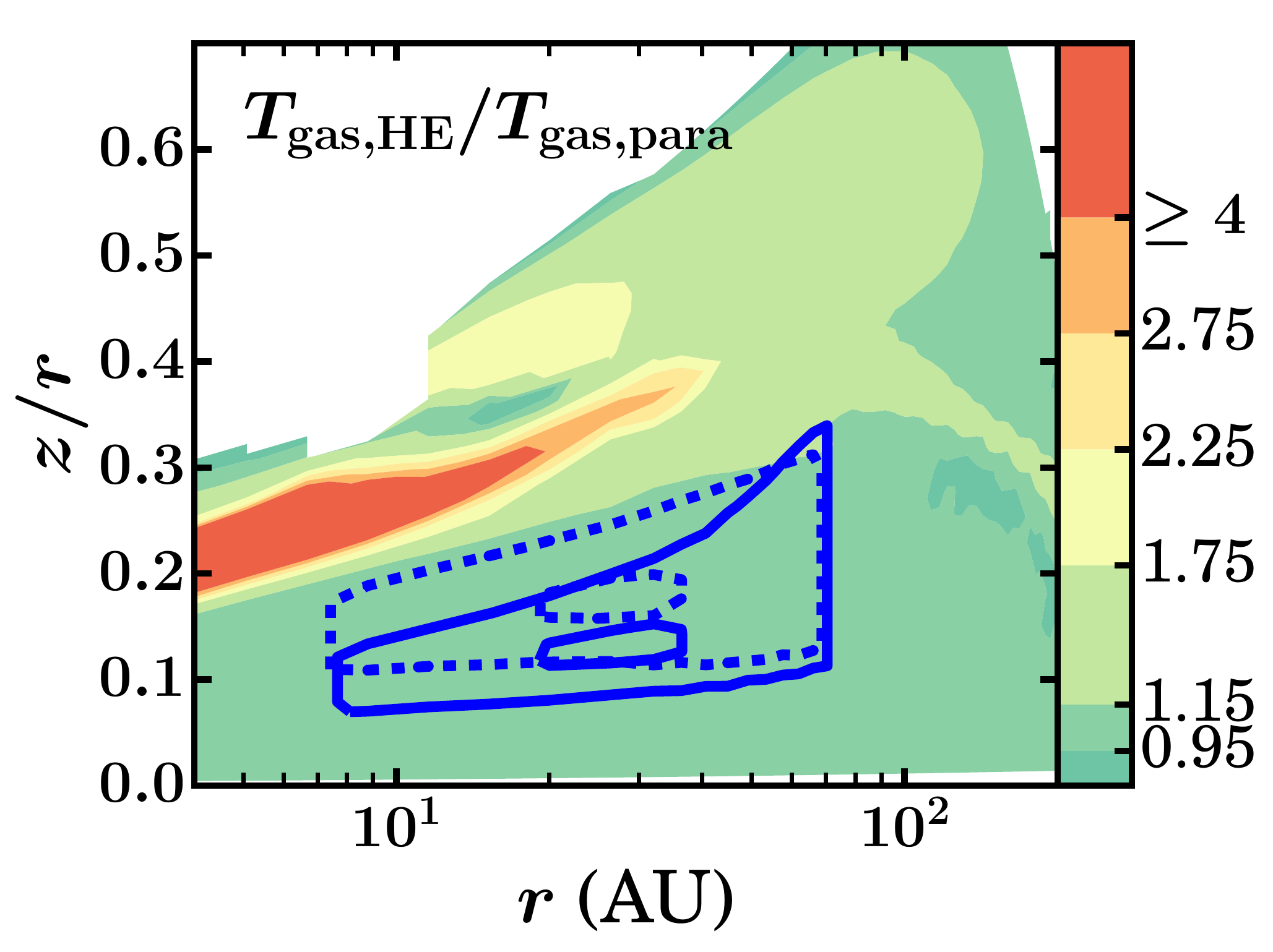}
    \end{subfigure}
    \caption{\label{fig: comparing HE} \newtext{Maps of the density (top) and temperature (bottom) differences between the fiducial model (denoted `para'; Section \ref{sec: HD emission maps}) and the same model after eight iterations of the hydrostatic solver (denoted `HE'). Colour indicates the ratio of the densities or gas temperatures of the two models. The dashed blue lines show the HD 1-0 emitting region of the fiducial model. The solid blue lines show the HD 1-0 the 25\% and 75\% cumulative emission regions of the `HE' model.}}
\end{figure}

\newtext{Figure \ref{fig: comparing HE} shows the effect of eight iterations of the hydrostatic solver on the density and gas temperature structure of the fiducial model (M$_{\rm disk} = 2.3\cdot10^{-2}\ \mathrm{M}_{\odot}$, $\psi = 0.3$ and $h_c = 0.1$, cf. Section \ref{sec: HD emission maps}). It also shows the regions where the HD 1-0 emission is produced before and after applying the hydrostatic solver. The top panel shows that the HD 1-0 emission has moved slightly towards the midplane for the hydrostatic model. This is likely the result of the decrease in height of the line optical depth surface. The bottom panel shows that the gas temperature in the HD 1-0 emitting region has changed by less than 5\%.}

\begin{figure}
\centering
\begin{subfigure}{\columnwidth}
\includegraphics[width=\columnwidth]{HD_distribution.pdf}
\end{subfigure}
\begin{subfigure}{\columnwidth}
\includegraphics[width=\columnwidth]{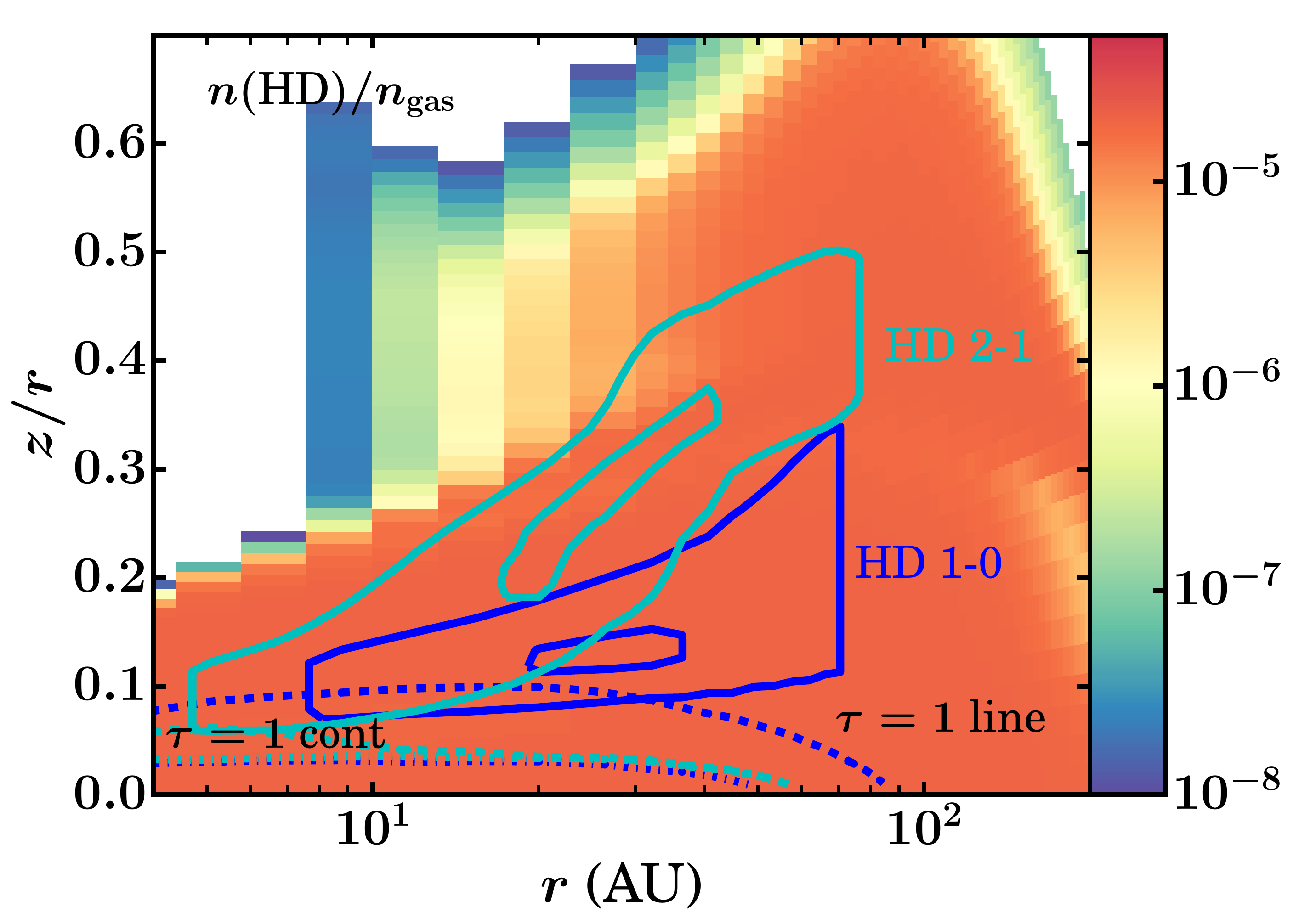}
\end{subfigure}
\caption{\label{fig: HDdistributionHydro}\textbf{Top panel:} \newtext{HD abundance structure for a disk model with M$_{\rm disk} = 2.3\cdot10^{-2}\ \mathrm{M}_{\odot}$, $\psi = 0.3$ and $h_c = 0.1$, identical to Figure \ref{fig: HDdistribution}. Colour indicates the number density of HD with respect to the total gas density. The coloured lines correspond the HD 1-0 (blue) and HD 2-1 (light blue) respectively. Solid contours indicate where 25\% and 75\% of the emission originates from. The dashed (dashed-dotted) lines show the $\tau=1$ surface for the line and continuum opacity, respectively.}
\textbf{Bottom panel:} \newtext{HD abundance structure of the same initial model, where the vertical structure is now determined by eight iterations of the hydrostatic solver of DALI.} }
\end{figure}

\newtext{The effect of the hydrostatic solver on the HD emitting regions is shown in Figure \ref{fig: HDdistributionHydro}. While disk models obtained with the hydrostatic calculations are vertically more extended, the HD emitting regions have not extremely changed. This is also reflected in the integrated fluxes, which for the hydrostatic model are $\sim50 \%$ (HD 1-0) and $\sim 66 \%$ (HD 2-1) lower compared to the model with the parametrized vertical structure. While this affects the inferred disk masses by a similar absolute factor, all the trends presented here should be robust. Also note that varying the parametric vertical structure changes the HD fluxes by a similar or larger factor (cf. Figure \ref{fig: fluxVS}). The mass uncertainties calculated in Section \ref{sec: determining the disk gas mass} therefore remain unchanged. }


\section{Deuterium chemistry}
\label{app: deuterium table}
Table \ref{tab: deuterium chemistry} list the reactions that were included in the deuterium chemical network. The rate coefficients were adapted from \cite{RobertsMillar2000,Walmsley2004} and \cite{GloverJappsen2007}. Where the rate coefficients  were not known, the coefficients for the analogue reactions with H instead of D were used.

\begin{table*}
\centering
\caption{\label{tab: deuterium chemistry}Simple HD chemical network reactions and adopted rate coefficient.}
\begin{tabular}{lrrrrl}
\hline\hline
Reaction & $\alpha$ & $\beta$ & $\gamma$ & Ref. \\
\hline
D$^+$ + H $\rightarrow $ H$^+$ + D & 1.0(-9) & 0.0(0) & 0.0(0)  & 1,2\\
D + H$^+$ $\rightarrow $ H + D$^+$ & 1.0(-9) & 0.0(0)  & 4.1(1) & 1,2\\
D$^+$ + H$_2$ $\rightarrow $ H$^+$ + HD & 2.1(-9) & 0.0(0)  & 0.0(0)  & 1,2\\
D$^+$ + H$_2$ $\rightarrow $ H + HD$^+$ & 1.75(-9) & 9.8(-2) & 0.0(0)  & 1,2\\
HD + H$^+$ $\rightarrow $ H$_2$ + D$^+$ & 1.0(-9) & 0.0(0)  & 6.345(2) & 1,2\\
HD$^+$ + H $\rightarrow $ HD + H$^+$ & 6.4(-10) & 0.0(0) & 0.0(0)& 2\\
H$_2^+$ + D $\rightarrow $ H$_2$ + D$^+$ & 6.4(-10) & 0.0(0) & 0.0(0)& 2\\
D + cr $\rightarrow $ D$^+$ + e$^-$ & 4.6(-20) & 0.0(0) & 0.0(0)& 2\\
HD + cr $\rightarrow $ H + D$^+$ + e$^-$ & 4.73(-20) & 0.0(0) & 0.0(0)& 2\\
HD + cr $\rightarrow $ D + H$^+$ + e$^-$ & 4.73(-20) & 0.0(0) & 0.0(0)& 2\\
HD + cr $\rightarrow $ D + H & 5.99(-19) & 0.0(0) & 0.0(0)& 2\\
HD + cr $\rightarrow $ D$^{+}$ + H$^-$ & 8.4(-22) & 0.0(0) & 0.0(0)& 2\\
HD + cr $\rightarrow $ HD$^+$ + e$^-$ & 5.16(-18) & 0.0(0) & 0.0(0)& 2\\
HD$^+$ + H$_2$ $\rightarrow $ H$_3^+$ + D & 5.25(-9) & 0.0(0) & 0.0(0)& 2\\
H$_2^+$ + HD $\rightarrow $ H$_3^+$ + D & 5.25(-9) & 0.0(0) & 0.0(0)& 2\\
HD + He$^+$ $\rightarrow $ He + H$^+$ +D & 5.5(-14) & -2.4(-1) & 0.0(0)& 2\\
HD + He$^+$ $\rightarrow $ He + D$^+$ +H & 5.5(-14) & -2.4(-1) & 0.0(0)& 2\\
D$^+$ + H $\rightarrow $ HD$^+$ + $\gamma$ & 3.9(-19) & 1.8(0) & 0.0(0)  & 2\\
H$^+$ + D $\rightarrow $ HD$^+$ + $\gamma$ & 3.9(-19) & 1.8(0) & 0.0(0)  & 2\\
HD$^+$ + H $\rightarrow $ H$_2^+$ + D & 1.0(-9) & 0.0(0) & 1.54(2)  & 2\\
H$_2^+$ + D $\rightarrow $ HD$^+$ + H & 1.0(-9) & 0.0(0) & 0.0(0)  & 2\\
HD$^+$ + D $\rightarrow $ HD$^+$ + D & 6.4(-10) & 0.0(0) & 0.0(0)  & 2\\
HD$^+$ + e$^-$ $\rightarrow $ H + D & 3.4(-9) &-4.0(-1) & 0.0(0)  & 2\\
D$^+$ + e$^-$ $\rightarrow $ D + $\gamma$ & 3.61(-12) &-7.5(-1) & 0.0(0)  & 2\\
H + D $\rightarrow $ HD + $\gamma$ & 3.0(-18) &5.0(-1) & 0.0(0)  & $^{*}$\\
D$^+$ + Mg $\rightarrow $ D + Mg$^+$ & 1.1(-9) & 0.0(0) & 0.0(0)  & $^{*}$\\
D$^+$ + Si $\rightarrow $ D + Si$^+$ & 9.9(-10) & 0.0(0) & 0.0(0)  & $^{*}$\\
D$^+$ + Fe $\rightarrow $ D + Fe$^+$ & 7.4(-9) & 0.0(0) & 0.0(0)  & $^{*}$\\
D$^+$ + S $\rightarrow $ D + S$^+$ & 1.3(-9) & 0.0(0) & 0.0(0)  & $^{*}$\\
D$^+$ + PAH $\rightarrow $ D + PAH$^+$ & 3.1(-8) & 0.0(0) & 0.0(0)  & $^{*}$\\
D$^+$ + PAH$^-$ $\rightarrow $ D + PAH & 8.3(-7) & -5.0(-1) & 0.0(0)  & $^{*}$\\
HD$^+$ + PAH $\rightarrow $ HD + PAH$^+$ & 3.1(-8) & 0.0(0) & 0.0(0)  & $^{*}$\\
HD$^+$ + PAH$^-$ $\rightarrow $ HD + PAH & 8.3(-7) & -5.0(-1) & 0.0(0)  & $^{*}$\\
\hline
\end{tabular}
 \captionsetup{width=0.54\textwidth}
 \caption*{\footnotesize{References: 1. \cite{RobertsMillar2000}; 2. \cite{Walmsley2004};\\ $^{*}$ Rate coefficients are the same as those for the analogue reactions with H instead of D.}}
\end{table*}


\section{Line fluxes of the TW Hya model}
\label{app: line fluxes TW Hya model}
The extended version of Figure \ref{fig: LineFluxes} is shown here. Its setup is similar to Figure 6 in \cite{Kama2016}. The fluxes and upper limits for the observations used can be found in Table B.1 in \cite{Kama2016}. The line fluxes of $^{13}$CO 6-5, C$^{18}$O 3-2 and C$^{18}$O 6-5 were calculated from the data presented in \cite{Schwarz2016}. 

\begin{figure*}
\centering
\includegraphics[width =  0.9\textwidth]{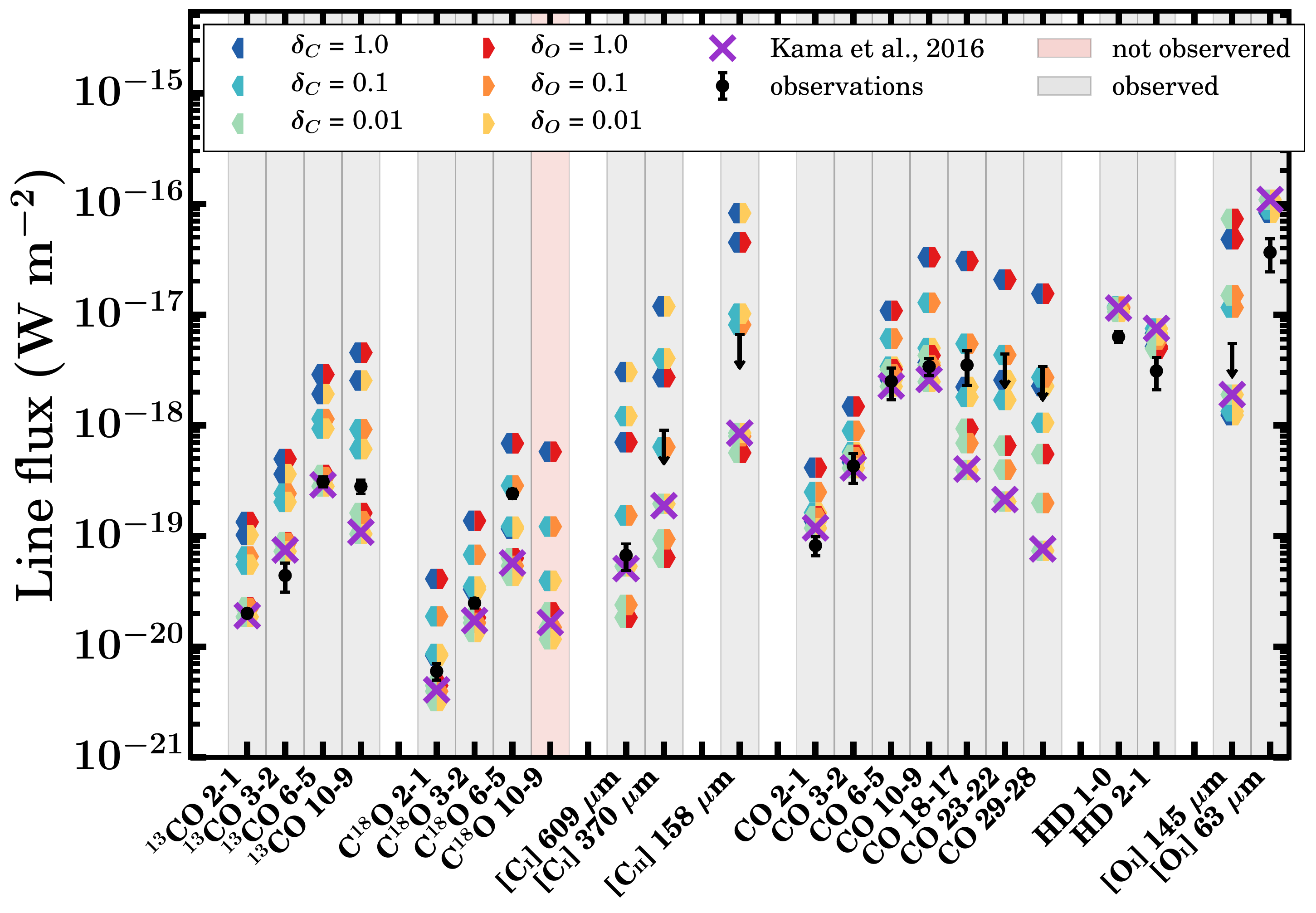}
\caption{\label{fig: LineFluxes Fullversion} 
Integrated line fluxes from TW Hya models (M$_{\rm disk} = 2.3\cdot10^{-2}\ \mathrm{M}_{\odot}$, $\psi = 0.3$, $h_c = 0.1$, , $d$ = 59.54 pc \citep{AstraatmadjaBailerJones2016}) with different amounts of carbon/oxygen underabundances. The x-axis shows the various modelled transitions. The black bars show the observations for TW Hya (\citealt{Schwarz2016,Kama2016} and references therein). The purple crosses show the modelled line fluxes from the best fit model from \cite{Kama2016} (with $\delta_C = \delta_O = 0.01$). The coloured hexagons show the results from this work, with the left hand side showing the amount carbon underabundance (with respect to the ISM). The right hand side shows the amount of oxygen underabundance. 
}
\end{figure*}

\end{appendix}

%
%

\end{document}